\documentclass[amsmath,amssymb,aps,twocolumn,superscriptaddress,showkeys,citeautoscript]{revtex4-2}
\pdfoutput=1
\usepackage{graphicx}
\usepackage{xcolor}
\usepackage[colorlinks,linkcolor=blue,citecolor=blue,urlcolor=blue]{hyperref}
\usepackage{amssymb}
\usepackage{comment}
\usepackage{xparse}
\usepackage{ifthen}

\def\tE{^{3}\!E}
\def\tA2{^{3}\!A_2}
\def\dE{^{2}\!E}
\def\qA2{^{4}\!A_2}
\def\dA2{^{2}\!A_2}
\def\sA1{^{1}\!A_1}
\def\sE{^{1}\!E}
\def\IPtA2{\mathrm{IP} (\tA2)}

\def\NVm{\rm{NV}^-}
\def\NV0{\rm{NV}^0}

\newcommand{\ket}[1]{
  \mathchoice%
  {\left|#1\right\rangle}       
  {|#1\rangle}                  
  {|#1\rangle}                  
  {|#1\rangle}                  
}

\newcommand{\matrixel}[3]{
  \mathchoice%
  {\left\langle#1 \middle| #2 \middle| #3 \right\rangle} 
  {\langle#1 | #2 | #3 \rangle}                   
  {\langle#1 | #2 | #3 \rangle}                   
  {\langle#1 | #2 | #3 \rangle}                   
}

\makeatletter

\newcommand*{\balancecolsandclearpage}{%
  \close@column@grid
  \cleardoublepage
  \twocolumngrid
}
\makeatother

\newboolean{draft}   
\setboolean{draft}{true}

\NewDocumentCommand\change{om}%
{%
  \ifthenelse{\boolean{draft}}{
    \IfNoValueTF{#1}{}%
    {%
      {\color{gray}[#1]}
    }%
    {\color{orange}#2}%
  }%
  {#2}%
}
\NewDocumentCommand\todo{m}%
{%
  {\color{blue} (TODO: #1)}%
}

\begin{document}


\title[NV photoionization]{Photoionization of negatively charged NV centers in diamond:\\
  theory and {\it ab initio} calculations}

\author{Lukas Razinkovas}
\email{lukas.razinkovas@ftmc.lt}
\affiliation{Center for Physical Sciences and Technology (FTMC),
  Vilnius LT--10257, Lithuania}

\author{Marek Maciaszek} 
\affiliation{Center for Physical Sciences and
  Technology (FTMC), Vilnius LT--10257, Lithuania}
\affiliation{Faculty of Physics, Warsaw University of Technology,
  Koszykowa 75, 00--662 Warsaw, Poland}

\author{Friedemann Reinhard}
\affiliation{Institute of Physics, University of Rostock, 18059
  Rostock, Germany}

\author{Marcus W. Doherty}
\affiliation{Laser Physics Centre, Research School of Physics, 
  Australian National University, Australian Capital Territory 2601, Australia}

\author{Audrius Alkauskas}
\email{audrius.alkauskas@ftmc.lt}
\affiliation{Center for Physical Sciences and Technology (FTMC),
  Vilnius LT--10257, Lithuania}
\affiliation{Department of Physics, Kaunas University of Technology
  (KTU), Kaunas LT--51368, Lithuania}


\date{\today}


\begin{abstract}
  We present {\it ab-initio\/} calculations of photoionization
  thresholds and cross sections of the negatively charged
  nitrogen--vacancy (NV) center in diamond from the ground $\tA2$ and
  the excited $\tE$ states. We show that after the ionization from the
  $\tE$ level the NV center transitions into the metastable $\qA2$
  electronic state of the neutral defect. We reveal how spin
  polarization of $\NVm$ gives rise to spin polarization of the $\qA2$
  state, providing an explanation of electron spin resonance
  experiments.  We obtain smooth photoionization cross sections by
  employing dense $k$-point meshes for the Brillouin zone integration
  together with the band unfolding technique to rectify the
  distortions of the band structure induced by artificial periodicity
  of the supercell approach. Our calculations provide a comprehensive
  picture of photoionization mechanisms of $\NVm$.  They will be
  useful in interpreting and designing experiments on charge-state
  dynamics at NV centers.  In particular, we offer a consistent
  explanation of recent results of spin-to-charge conversion of NV
  centers.
\end{abstract}

\maketitle


\section{Introduction\label{sec:intro}}

Over the past two decades the nitrogen-vacancy (NV) center in
diamond~\cite{doherty2013} has become one of the key
platforms~\cite{awschalom2018} to test and eventually implement
various quantum technologies. Most technology-ready applications have
been in the field of quantum sensing~\cite{schirhagl2014}, but
progress in quantum communication~\cite{hensen2015} and quantum
computing~\cite{bradley2019,pezzagna2021} has been eminent, too. The
spin of the negatively charged NV center can be polarized and read out
optically~\cite{jelezko2004}. It has been established that the use of
optical excitation can lead to the photoionization of $\NVm$, whereby
an electron from the NV center is excited to the conduction band and
$\NVm$ is converted to $\NV0$~\cite{beha2012,siyushev2013,aslam2013}.
In~many situations this is a detrimental process for the operation of
$\NVm$ and it has to be avoided by carefully choosing experimental
parameters. Photoionization is also disadvantageous for the operation
of diamond lasers based on NV centers~\cite{jeske2017}, as it competes
with stimulated emission. To better differentiate between the two
mechanisms, the knowledge of the photoionization threshold and cross
section from the $\tE$ state would be very helpful.

In cases discussed above photoionization of $\NVm$ is
detrimental. However, {\it deliberate\/} photoionization of $\NVm$ can
be also very beneficial. In particular, it has been used to develop
the so-called photocurrent detection of magnetic resonance
(PDMR)~\cite{bourgeois2015}.  This technique can in principle reach
spin read-out rates superior to optical protocols, as the latter are
ultimately limited by radiative lifetimes. PDMR imaging of a single NV
center has been achieved~\cite{siyushev2019} and, very recently, the
detection of a single nuclear spin by PDMR has been
demonstrated~\cite{gulka2021}.  Photoionization of NV centers is also
used for spin read-out via spin-to-charge
conversion~\cite{waldherr2011,shields2015,hopper2018}.  Lastly,
excitation to the conduction band plays an important role in a
proposed protocol to couple two remote NV centers using spatial
stimulated Raman adiabatic passage~\cite{oberg2019}.

\begin{figure}
\begin{center}
\includegraphics[width=1\linewidth]{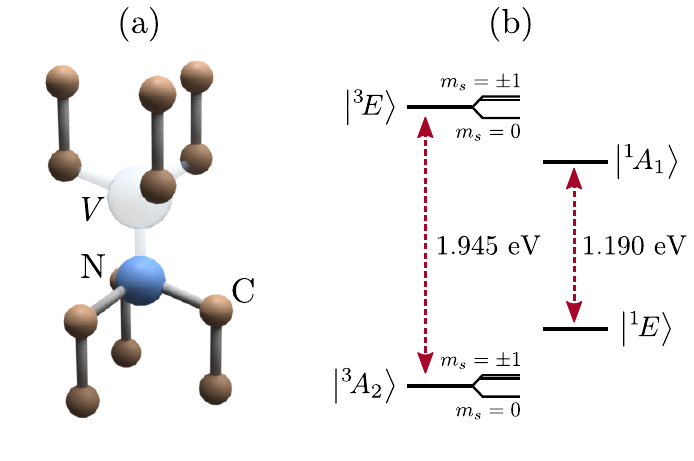}
\caption{(a) Atomic structure of the nitrogen-vacancy center in
  diamond. (b) Electronic level diagram of $\NVm$. Energies of the ZPL
  between the triplets and the singlets are indicated. Spin sublevels
  of $\tA2$ and $\tE$ states are not to scale.\label{fig:levels0}}
\end{center}
\end{figure}

The atomic structure of the NV center and the energy-level diagram of
its negative charge state are shown in Fig.~\ref{fig:levels0}.  There
are four electronic levels: spin-triplet states $\tA2$ and $\tE$, as
well as metastable spin singlets $\sE$ and $\sA1$.  It has been
established that photoionization of $\NVm$ can occur either via a
one-photon or a two-photon
mechanism~\cite{beha2012,siyushev2013,aslam2013}.  In a single-photon
ionization an electron from the $\tA2$ ground state is directly
promoted to the conduction band. The threshold for the process has
been experimentally determined to
be~${\sim}2.6$~eV~\cite{aslam2013}. The second mechanism is a
sequential process of a two-photon
absorption~\mbox{\cite{beha2012,siyushev2013,aslam2013}}.  In~this
case the NV center is first excited to the $\tE$ state; the
zero-phonon line (ZPL) of this transition is
$E_{\textrm{ZPL}}=1.945$~eV~\cite{doherty2013}. Subsequently, the NV
center is ionized from the $\tE$ state. In most practical situations,
both when photoionization is beneficial or detrimental, it is the
latter process that is most
important~\cite{beha2012,siyushev2013,aslam2013}.  These two processes
do not exhaust all the possibilities. After the absorption of the
first photon the NV center can undergo an inter-system crossing (ISC)
to the singlet $\sA1$ level
(Fig.~\ref{fig:levels0})~\cite{doherty2013}.  This is a short-lived
state with a lifetime of $0.1$~ns~\cite{Ulbricht2018}, from which
there is a mostly nonradiative transition to the $\sE$ state. The
latter is long-lived with a lifetime
$150\mbox{--}450$~ns~\cite{Acosta2010}, enabling photoionization from
this state via the absorption of the second photon.  This is the third
photoionization mechanism. Photoionization from the singlets was
invoked previously~\cite{bassett2016}, but the mechanism of the
process was not investigated in detail in literature.

Experimental measurements of photoionization cross sections and
thresholds for the NV center are not straightforward, in particular
regarding the photoionization from the excited state $\tE$. The first
difficulty is related to the fact that light can induce both the
transition $\NVm\to\NV0$ (ionization) and $\NV0\to\NVm$
(recombination), often making it hard to disentangle the two
processes~\cite{beha2012,siyushev2013,aslam2013}.  Moreover, for the
NV center in the excited triplet state photoionization competes with
stimulated emission, whereby the NV center returns back to the ground
state~\cite{jeske2017,jeske2020}, marring the experimental handle of
the photoionization process even further. To the best of our
knowledge, neither absolute photoionization cross sections from
$\tA2$, $\tE$, and $\sE$ states, nor photoionization thresholds from
$\tE$ and $\sE$ states have yet been determined experimentally.

In this paper we address the question of the photoionization
thresholds and absolute photoionization cross sections using {\it
  ab-initio\/} calculations. In the two-photon ionization the
intra-defect absorption precedes the ionization step. Therefore, we
calculate the cross section for that process as well.  We also report
calculations for the cross section of the stimulated emission from the
$\tE$ state. Our work provides important new knowledge about
ionization mechanisms that differs from earlier
work~\cite{siyushev2013}. The main focus of our work is
photoionization from triplet states $\tA2$ and $\tE$.  However, we
also discuss photoionization from the $\sE$ state.  {\it Ab initio}
treatment of the $\sE$ state is more complicated \cite{Gali2008} and
we resort to a semi-quantitative analysis based on available
experimental data and calculations for the other two states.

This paper is organized as follows. In Sec.~\ref{sec:mechanism} we
discuss the mechanism of photoionization of $\NVm$ centers in more
detail.  We give the expressions for photoionization thresholds and
cross sections, as well as cross sections for intra-defect absorption
and stimulated emission. In Sec.~\ref{sec:methods} we discuss the
particulars of the electronic structure, introduce computational
methods and approximations to calculate photoionization thresholds and
cross sections, as well as cross sections for intra-defect absorption
and stimulated emission. We present the results of calculations and
their analysis in Sec.~\ref{sec:results}. The consequences of our work
to the physics of NV centers are discussed in
Sec.~\ref{sec:discussion}. Finally, Sec.~\ref{sec:conclusions}
concludes our work. The target groups of our paper are (i) broad
community working on the physics and applications of color centers and
(ii) theorists interested in the development of computational
methodologies for point defects in solids. The first group can skip a
rather technical Sec.~\ref{sec:methods}. In the paper we use
$\epsilon$ for photon energies, $E$ for electron energies.


\section{NV center photoionization mechanisms\label{sec:mechanism}}

\begin{figure}
\begin{center}
\includegraphics[width=1.0\linewidth]{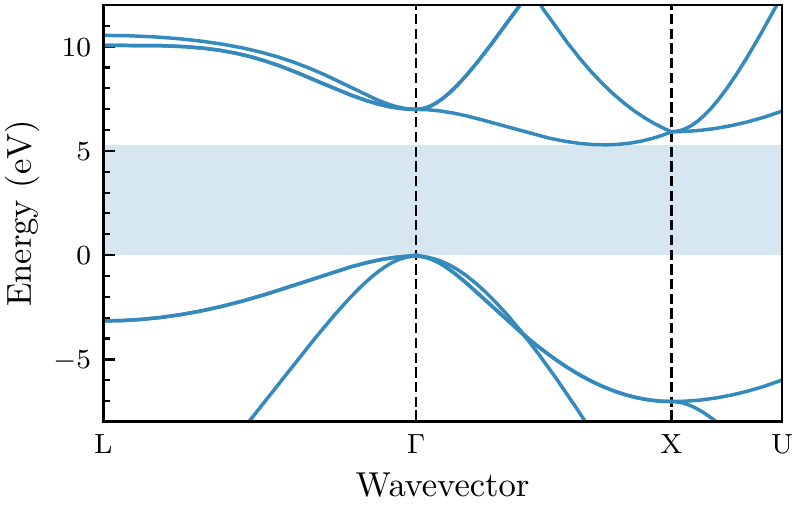}
\caption{Band structure of bulk diamond calculated using the HSE
  density functional (see Sec.~\ref{sec:methods} for more details about
  computational methods). The shaded area corresponds to the band gap.
  Photoionization threshold is determined by electrons being excited to 
  the CBM that occurs between $\Gamma$ and
  $X$ points.\label{fig:band}}
\end{center}
\end{figure}

\subsection{Photoionization thresholds\label{sec:threshold}}

A threshold for
photoionization corresponds to electron being excited to the
conduction band minimum (CBM). In the case of diamond the CBM occurs
along the $\Gamma - X$ line in the Brillouin zone, as shown in
Fig.~\ref{fig:band}. 

\subsubsection{Photoionization from the $\tA2$ state}

\begin{figure}
\begin{center}
\includegraphics[width=1.0\linewidth]{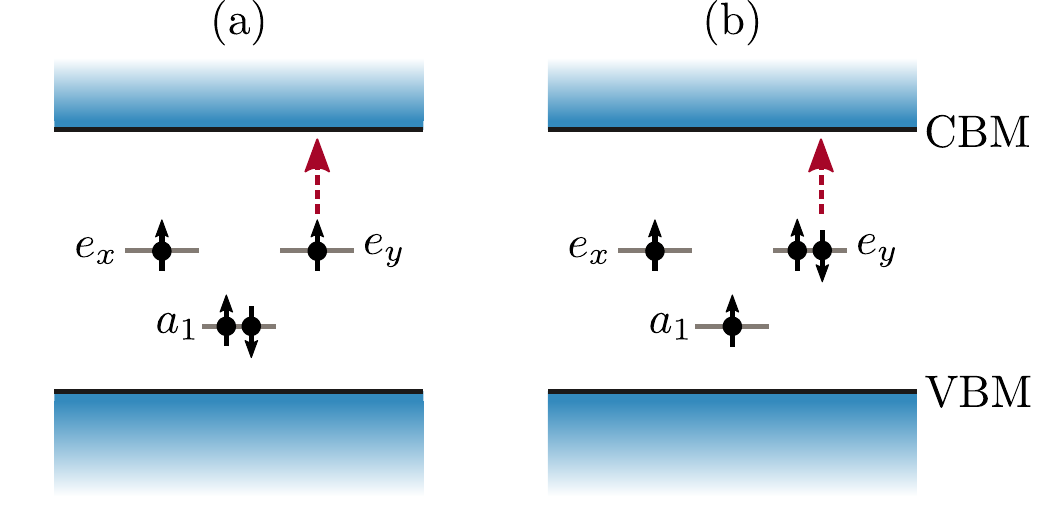}
\caption{Photoionization of $\NVm$ center in the single-electron
  picture.  (a) Electronic configuration of the $m_s=1$ spin sublevel
  of the $\tA2$ state.  (b) Electronic configuration of the $m_s=1$
  spin sublevel of the $E_x$ component of the $\tE$ manifold.  Red
  arrows show one possibility of photoionization, whereby an $e_y$
  electron is excited to the conduction band (see text for a more
  in-depth discussion).\label{fig:mech}}
\end{center}
\end{figure}

\begin{figure}
\begin{center}
\includegraphics[width=1.0\linewidth]{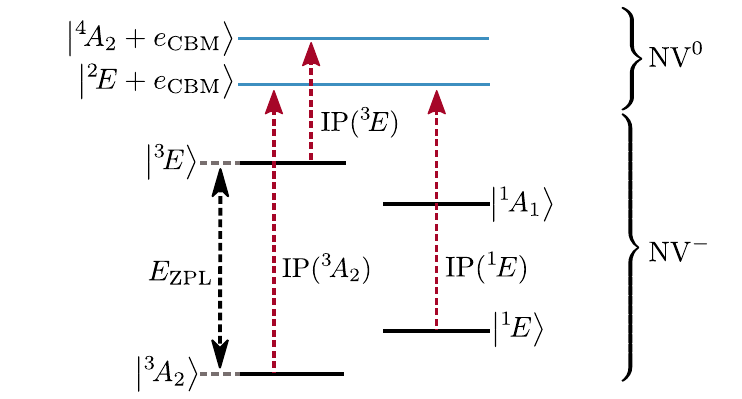}
\caption{Photoionization of $\NVm$ from $\tA2$, $\tE$, and $\sE$
  states.  Horizontal lines indicate the energy of the system: black
  for $\NVm$ and blue for $\NV0$ plus an electron at the CBM. All
  energies are to scale. Red arrows indicate possible photoionization
  mechanisms, $E_{\text{ZPL}}$ is the ZPL energy of the triplet
  transition.\label{fig:levels}}
\end{center}
\end{figure}

The ground state of $\NVm$, $\tA2$, is described by an electron
configuration $a_1^2e^2$, where $a_1$ and $e$ label irreducible
representations of single-particle levels~\cite{doherty2013}.
$m_s=\pm 1$ spin sublevels of the triplet manifold can be described as
single Slater determinants. The $m_s=1$ state is illustrated in
Fig.~\ref{fig:mech}(a); in the ket notation it can be written as
$\ket{a_1 \bar{a}_1 e_x e_y}$, where ``bar'' indicates spin-down
electrons. In the single-particle picture ionization is a process
whereby one electron from the $e$ level is excited to the conduction
band, turning $\NVm$ into $\NV0$.  This is illustrated by a red dotted
arrow in Fig.~\ref{fig:mech}(a).  After the $\NVm$ is ionized, NV
center transitions into the $\dE$ ground state of the neutral center
with electron configuration $a_1^2e^1$.  The photoionization process
can be depicted using the energy-level diagram of the {\it entire}
system, i.e.~$\NVm$ or $\NV0$ plus an electron at the CBM, as shown in
Fig.~\ref{fig:levels}. Photoionization threshold from the $\tA2$
state, $\IPtA2$, has been measured experimentally by Aslam {\it et
  al.\/}~\cite{aslam2013}. Careful study of charge conversion dynamics
as a function of wavelength and intensity of laser illumination
provided the value $\IPtA2=2.6$~eV; the error bar of this value can be
assumed to be ${\sim}0.1$~eV~\cite{aslam2013}.  Previous
state-of-the-art theoretical calculations yielded the result
$\textrm{IP} (\tA2)=2.64$~eV~\cite{deak2014}, in excellent agreement
with experiment.

\subsubsection{Photoionization from the $\tE$ state}

Compared to the photoionization from the ground state, the physics of
the photoionization from the $\tE$ state is presently less understood.
To the best of our knowledge, the photoionization threshold for
photoionization from the excited triplet state is not known. The
electronic configuration of the $\tE$ state is $a_1^1e^3$. The $m_s=1$
spin sublevel of the $E_x$ orbital component,
$\ket{a_1 e_x e_y \bar{e}_y}$, is illustrated in
Fig.~\ref{fig:mech}(b).  Removing one electron from the $e$ level [red
dotted arrow in Fig.~\ref{fig:mech}(b)] yields the configuration
$a_1^1e^2$.  The lowest-energy state with this configuration is the
spin-quartet $\qA2$ state of $\NV0$.  Therefore, the final state of
the $\NV0$ after the photoionization is the meta-stable state $\qA2$
and not the ground state $\dE$, as it was sometimes assumed
\cite{siyushev2013}.  The energy-level diagram for this
photoionization process is shown in Fig.~\ref{fig:levels}: the initial
state is $\NVm$ in the $\tE$ state, while the final state is $\NV0$ in
the $\qA2$ state plus an electron in the conduction band.  The
expression for the threshold of photoionization, $\textrm{IP} (\tE)$,
can be read from Fig.~\ref{fig:levels}:
\begin{equation}
\textrm{IP}(\tE) =
\textrm{IP}(\tA2) - E_{\textrm{ZPL}}
+ \left[E\left(\qA2\right) - E\left(\dE\right) \right].
\label{IP2}
\end{equation}

The equation above enables the determination of the threshold for the
photoionization from the $\tE$ state. Unfortunately, the energy
difference $\left[E\left(\qA2\right) - E\left(\dE\right) \right]$
between the two states of $\NV0$ is not known
experimentally. Therefore, the ``experimental'' value of
$\textrm{IP}(\tE)$ cannot be deduced from this relationship. Even
though the experimental values of $\textrm{IP}(\tA2)$ and
$E_{\textrm{ZPL}}$ are known, to benefit from possible cancellation of
errors in theoretical calculations, in this paper we will calculate
all the quantities that appear in Eq.~\eqref{IP2} using the same
computational setup, described in Sec.~\ref{sec:methods}.

Apart from the $\qA2$ state, there are other $\NV0$ states with the
electron configuration $a_1 e^2$, the lowest of which is
$\dA2$~\cite{doherty2013}. Using the equation similar to
Eq.~\eqref{IP2} we can determine the ``experimental'' threshold for
the photoionization via this process to be about 2.8~eV. This is
outside the range of energies we consider in this work and therefore
this process will be not be analyzed further.

\subsubsection{Photoionization from the $\sE$ state\label{sec:sE}}

Unlike $m_s=\pm 1$ spin sublevels of the triplet states, the two
components of the orbital doubled $\sE$ are described by
multi-determinant wavefunctions, as discussed in, e.g.,
Ref.~\cite{Gali2008}.  However, as in the case of $\tA2$, the
electronic configuration of the $\sE$ state is $a_1^2e^2$.  During the
photoionization one $e$ electron is promoted to the conduction band,
and NV center transitions to the $\dE$ ground state of the neutral
defect with electronic configuration $a_1^2e^1$.  It is
straightforward to deduce the photoionization threshold from the $\sE$
state (Fig.~\ref{fig:levels}):
\begin{equation}
\textrm{IP}(\sE) =
\textrm{IP}(\tA2) - \left[E\left(\sE\right) - E\left(\tA2\right) \right].
\label{IP3}
\end{equation}
The energy difference
$\left[E\left(\sE\right) - E\left(\tA2\right) \right]$ has not been
measured directly.  However, the analysis of the ISC between the $\tE$
and the $\sA1$, as well as the knowledge of the ZPL energy between the
two singlets, enables one to determine this energy difference to be
about $0.38$~eV~\cite{goldman2015}. As a result, $\textrm{IP}(\sE)$
can be estimated to be $2.2 \pm 0.1$ eV.

\subsection{Photoionization cross sections\label{sec:cross sections}}

The general theory of optical absorption in semiconductors is given in
a number of textbooks, e.g., Refs.~\cite{ridley,stoneham}.  Let
$\tilde{\sigma}_{\rm{ph}}(\epsilon)$ be the photoionization cross
section as a function of photon energy $\epsilon$ in the absence of
lattice relaxation.  It is given by (cf.~Eq.~(10.2.4) in
Ref.~\cite{stoneham}):
\begin{align}
  \tilde{\sigma}_{\rm{ph}}(\epsilon) =
  &
    \frac{4\pi^2 \alpha}{3n_D} 
    \epsilon
    \sum_{j}
    r_{ij}^2
    \delta
    \left(
    \epsilon - E_{ij}
    \right).
\label{sigma1}
\end{align}
Here $\alpha$ is the fine-structure constant and $n_{D}=2.4$ is the
refractive index of diamond. Label $i$ denotes the initial state
$\Psi_i$ and the sum runs over all final states $\Psi_j$;
$E_{ij}=E_j - E_i$ is the energy difference between the two
states. $\vec{r}_{ij}$ are transition dipole moments (that we will
also call optical matrix elements), discussed in
Sec.~\ref{sec:elements}.  We consider the absorption light by an
ensemble of randomly oriented NV centers.  This is the reason for the
appearance of the factor $1/3$ in Eq.~\eqref{sigma1}.

Vibrational broadening is introduced by replacing $\delta(\epsilon)$
in Eq.~\eqref{sigma1} with normalized spectral functions of
electron--phonon coupling $A(\epsilon)$ (for a more thorough
discussion see, e.g., Refs.~\cite{alkauskasNJP2014,razinkovas2020}).
In this case we can write actual cross section as a convolution:
\begin{align}
  \sigma_{\rm{ph}}(\epsilon)
  & = \epsilon\int_{-\infty}^{\infty}
     \frac{1}{\epsilon'}\tilde{\sigma}_{\rm{ph}}(\epsilon')
    A(\epsilon-\epsilon')\,\mathrm{d}\epsilon'.
\label{sigma2}
\end{align}
Temperature dependence of the photoionization cross section occurs
mainly via the temperature dependence of the spectral functions
$A(\epsilon)$.  In the remainder of the paper we will assume a $T=0$~K
limit for these functions.

\subsection{Cross section of the intra-defect absorption and
  stimulated emission\label{sec:stimulated}}

As discussed in Sec.~\ref{sec:intro}, photoionization from the $\tE$
state competes with stimulated emission that returns $\NVm$ back to
the ground state $\tA2$. The cross section of stimulated emission (the
process $\tE \rightarrow {\tA2}$) is given by an expression:
\begin{equation}
  \sigma_{\rm{st}}(\epsilon)
  =
\frac{4\pi^2 \alpha}{3n_D}  \epsilon r_{ij}^2 A \left (E_{\mathrm{ZPL}} - \epsilon \right).
\label{stimulated}
\end{equation}
Here $r_{ij}$ is the optical matrix element for the transition
${\tA2}\rightarrow{\tE}$, $E_{\rm{ZPL}}=1.945$~eV, and $A(\epsilon)$
is the spectral function for electron--phonon coupling for stimulated
emission, which is identical to that of spontaneous emission
(luminescence).

Another important parameter needed to understand the whole two-step
photoionization process is the cross section for intra-defect
absorption $\tA2 \to {\tE}$. Its cross section is given by:
\begin{equation}
  \sigma_{\rm{intra}}(\epsilon)
  =
  g \frac{4\pi^2 \alpha}{3n_D}  \epsilon r_{ij}^2 A
  \left(
    \epsilon - E_{\mathrm{ZPL}}
  \right).
\label{intra}
\end{equation}
Here $A(\epsilon)$ is spectral function of electron--phonon coupling
for the absorption $\tA2 \to {\tE}$, and $g = 2$ is the orbital
degeneracy factor of the final state $\tE$.


\section{Theory and methods}\label{sec:methods}

\subsection{Electronic structure methods}

Calculations have been performed within the framework of density
functional theory (DFT). For the geometry optimization, as well as the
calculation of excitation energies and ionization thresholds we used
the hybrid exchange and correlation functional of Heyd, Scuseria, and
Ernzerhof (HSE)~\cite{heyd2003}. In this functional a fraction $a=1/4$
of screened Fock exchange is ad-mixed to the semi-local exchange based
on the generalized gradient approximation in the form of Perdew,
Burke, and Ernzerhof~\cite{perdew1996}. As discussed below, in order
to obtain converged photoionization cross sections, we have to perform
integration on a dense $k$-point grid in the Brillouin
zone. Unfortunately, such calculations are computationally too
expensive if done with the HSE functional. For this purpose the
optical matrix elements have been calculated using the PBE
functional. Calculations for selected transitions have shown that HSE
and PBE matrix elements differ by less than~10\%. We used the
projector-augmented wave approach with a plane-wave energy cutoff of
500~eV. Calculations have been performed with the Vienna Ab-initio
Simulation Package ({\sc Vasp})~\cite{vasp}.

HSE functional provides a very good description of diamond, yielding
the band gap of $E_g=5.34$~eV (experimental value $5.46$~eV), and the
lattice constant $a=3.548$~\AA\ (experimental value
$3.567$~\AA). Geometry relaxation of the NV center has been performed
using $4 \times 4 \times 4$ supercells~\cite{freysoldt2014} with 512
atomic sites and a single $\Gamma$ point for the Brillouin zone
sampling. Ionization potential from the ground state
$\textrm{IP}(\tA2)$ has been determined from computed charge-state
transition levels as discussed in, e.g., Ref.~\cite{freysoldt2014}
with finite-size electrostatic corrections of
Ref.~\cite{freysoldt2009}. Spectral functions of electron--phonon
coupling $A(\epsilon)$ that appear in Eq.~\eqref{sigma2} have been
calculated following the methodology of Ref.~\cite{razinkovas2020}
(see the Sec.~I of the Supplemental Material~\cite{supp} for a more
detailed discussion). Spectral functions for absorption and stimulated
emission in Eqs.~\eqref{stimulated} and~\eqref{intra} have been taken
from Ref.~\cite{razinkovas2020}.

The energies of excited states $\tE$ and $\qA2$ that appear in
Eq.~\eqref{IP2} have been calculated using the
delta-self-consistent-field ($\Delta$SCF) method~\cite{jones1989},
first applied to the NV center by Gali {\it et
  al.\/}~\cite{gali2009}. To calculate the energy of the $\tE$ state
the spin-minority electron in the $a_1$ level is promoted to the $e$
level.  The total energy of the $\qA2$ state was calculated setting
the spin projection to $m_s=+3/2$.  The $\Delta$SCF method typically
performs very well when (i)~the state is described by single Slater
determinant and (ii)~the state has a different spin and/or orbital
symmetry from the ground state~\cite{jones1989}. This is indeed the
case for $\tE$ and $\qA2$ states in with spin projections $m_s=\pm 1$
and $m_s = \pm 3/2$, respectively.

\subsection{The nature of electronic states and calculations of
  optical matrix elements\label{sec:elements}}

The initial state $\Psi_i$ that enters in the calculation of the
matrix element $\vec{r}_{ij}$ in Eq.~\eqref{sigma1} represents the
entire solid with an embedded negatively charged defect. The final
state $\Psi_j$ represents the solid with a neutral defect plus an
excited electron in the conduction band. The optical matrix element
$\vec{r}_{ij}$ and energy difference $E_{ij}$ should then be
calculated for these many-electron states.  The calculation of the
matrix elements for multi-electron wavefunctions is a computationally
difficult problem and we will use approximations as described below.

\subsubsection{Photoionization from the $\tA2$ state}

\def\a{a_1}
\def\ab{\bar{a}_1}
\def\ex{{e}_x}
\def\ey{{e}_y}
\def\exb{\bar{e}_x}
\def\eyb{\bar{e}_y}

Let us first assume that $\NVm$ is initially in the \mbox{$m_s=1$}
spin sublevel. As already discussed above, this state can be described
by a single Slater determinant
$\ket{\tA2;1}=\ket{a_1\bar{a}_{1}e_{x}e_{y}}$
[Fig.~\ref{fig:mech}(a)]. The final state
$\ket{(\dE_{x/y};\tfrac{1}{2})\nobreak \otimes \nobreak
  \phi_{c}}=\ket{a_1\bar{a}_1e_{x/y}\phi_{c}}$ is an antisymmetrized
product of $\NV0$ in the $\dE$ state with spin sublevel
\mbox{$m_s=\tfrac{1}{2}$},
$\ket{\dE_{x/y}; \tfrac{1}{2}}=\ket{a_1\bar{a}_1e_{x/y}}$, and a
spin-up electron in the conduction band with a wavefunction $\phi_c$.

Let $\hat{O}=\sum_i\vec{r}_i$ be the many-electron dipole operator. To
simplify the calculation of matrix elements we will assume that all
single-electron orbitals from which many-electron wavefunctions are
formed are the same in the initial and the final state; the final
state differs from initial one by a single occupied orbital, which
corresponds to electron in the $e$ state being excited to the
conduction band. Such simplification allows to adopt the
Slater--Condon rule and reduce the matrix element calculated for a
many-body wavefunctions to a matrix element between the two Kohn-Sham
states:
\begin{equation}
  \vec{r}_{ij}
  \equiv
  \matrixel{\tA2;1}{\hat{O}}{(\dE_{x/y}; \tfrac{1}{2})
    \otimes \phi_c}
  = \matrixel{e_{y/x}}{\vec{r}}{\phi_c}.
\end{equation}
As the final state is an orbital doublet, we can calculate, for
example, only transition to the $E_x$ state and multiply the final
result by the degeneracy factor $g=2$.  The reasoning for the $m_s=-1$
sublevel is analogous, with the only difference that the spin-down
electron is excited to the conduction band.

The spin state $m_s=0$ is a sum of two Slater determinants
$\ket{\tA2;0}=1/\sqrt{2}(\ket{\a\ab\ex\eyb}+\ket{\a\ab\exb\ey})$.
In~this case the transition is possible to all four states of the
$\dE$ manifold, each with matrix element of the type
$\matrixel{e_{y/x}}{\vec{r}}{\phi_c}\sqrt{2}$.  The resulting overall
cross section is the same as for the $m_s=\pm 1$ spin projections.

\subsubsection{Photoionization from the $\tE$ state \label{sec:tEa}}

The situation with the photoionization from the $\tE$ is more
delicate.  Let us first consider the $m_s=1$ spin sublevel with a
wavefunction
$\ket{\tE_{x/y};1}=\ket{a_{1}e_{x}e_{y}\bar{e}_{y/x}}$. Now there are
two different channels for photoionization.  (i) A spin-down electron
is promoted to the conduction band [red arrow in
Fig.~\ref{fig:mech}(b)]. The final state is an antisymmetrized product
of $\ket{\qA2;\tfrac{3}{2}}$, describing $\NV0$ in the $\qA2$ manifold
with spin projection $m_s=3/2$, and a spin-down electron in the
conduction band:
$\ket{(\qA2;\tfrac{3}{2})\otimes\bar{\phi}_c}=\ket{a_{1}e_{x}e_{y}\bar{\phi}_{c}}$.
Since both states are single Slater determinants, using the
Slater--Condon rule we obtain:
\begin{equation}
  \vec{r}_{ij}
  \equiv
  \matrixel{\tE_{x/y};1}{\hat{O}}{(\qA2; \tfrac{3}{2}) \otimes \bar{\phi}_c}
  = \matrixel{\bar{e}_{y/x}}{\vec{r}}{\bar{\phi}_c}.
\end{equation}
(ii) A spin-up $e$ electron can also be excited to the conduction
band. In~this case the NV center transitions to the $m_s=1/2$ spin
sublevel of the $\qA2$ manifold, which is a multi-determinant state.
The final state of the entire system can be written as:
$\ket{(\qA2;\tfrac{1}{2})\otimes \phi_c}= 1/\sqrt{3}(%
\ket{\bar{a}_1e_{x}e_{y}\phi_{c}} + \ket{a_1\bar{e}_{x}e_{y}\phi_{c}}
+ \ket{a_1e_{x}\bar{e}_{y}\phi_{c}})$.  Applying the Slater--Condon
rule we obtain:
\begin{equation}
  \vec{r}_{ij}
  \equiv
  \matrixel{\tE_{x/y};1}{\hat{O}}{(\qA2; \tfrac{1}{2}) \otimes {\phi}_c}
  =
  \frac{1}{\sqrt{3}}\matrixel{{e}_{y/x}}{\vec{r}}{{\phi}_c}.
\end{equation}
As the photoionization rate depends on $r_{ij}^2$, we see that the
probability of the transition to the $m_s=1/2$ sublevel of the $\qA2$
manifold is 3 times smaller than to the $m_s=3/2$ sublevel (we will
ignore the difference between matrix elements calculated for spin-up
and spin-down Kohn-Sham states).  Formally, when evaluating the total
photoionization cross section pertaining to the $m_s=1$ sublevel of
the $\tE$ state via Eq.~\eqref{sigma1}, we can consider only the
excitation of the spin-down electron and multiply the final result by
the ``degeneracy factor'' $g=4/3$.  This is the procedure we have
chosen in this work. Analogous reasoning holds for the photoionization
from the $m_s=-1$ spin sublevel.

\begin{figure}
\begin{center}
\includegraphics[width=0.99\linewidth]{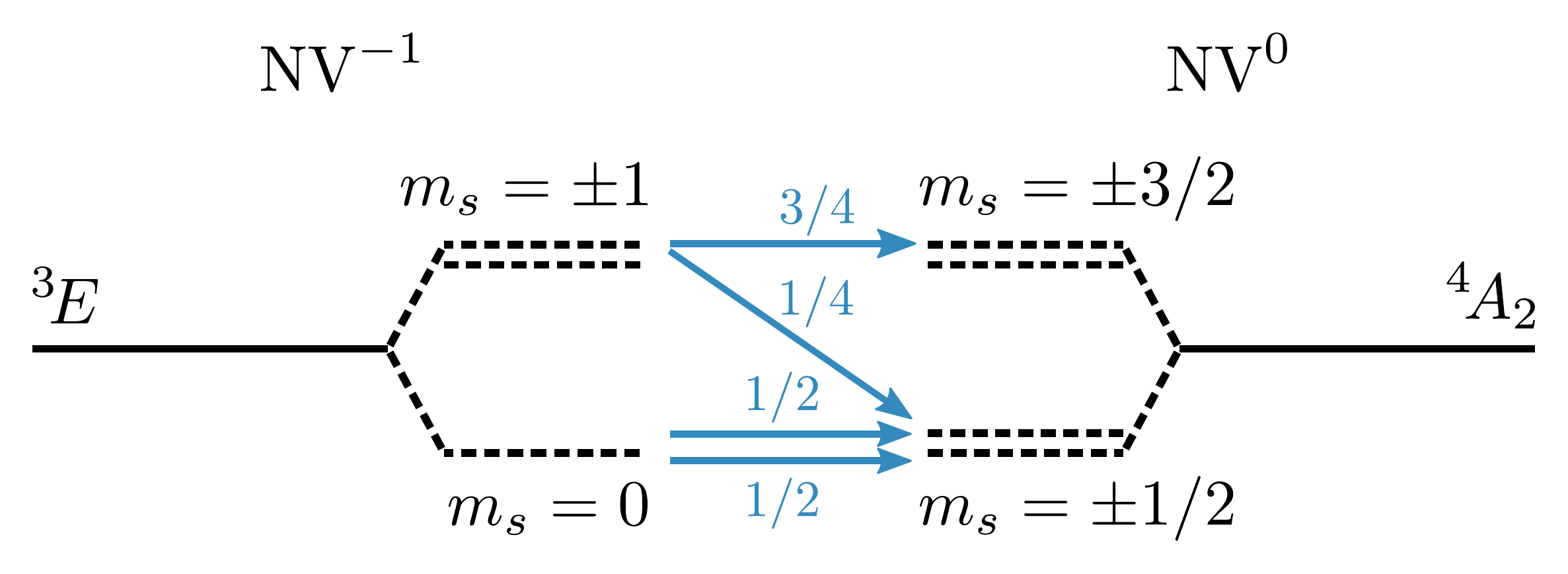}
\caption{Spin physics of the photoionization from the $\tE$ state.
  Numbers near arrows show relative probabilities of the transition
  during photoionization.  The $m_s=+1$ ($-1$) spin sublevel
  transitions into either the $m_s=+3/2$ ($-3/2$) or the $m_s=+1/2$
  ($-1/2$) sublevel of the $\qA2$ manifold with different
  probabilities.  The $m_s=0$ sublevel transitions to the
  $m_s=\pm 1/2$ sublevels with equal probability. Spin sublevels are
  separated by zero-field splittings
  $D(\tE) = 1.42$~GHz~\cite{doherty2013} and
  $D (\qA2) = 1.69$~GHz~\cite{felton2008}.\label{fig:qA2}}
\end{center}
\end{figure}

If $\NVm$ is initially in the $m_s=0$ sublevel
$\ket{\tE_{x/y};0}=1/\sqrt{2}(\ket{a_{1}\bar{e}_{x}e_{y/x}\bar{e}_{y/x}}
+ \ket{\bar{a}_1e_{x}e_{y/x}\bar{e}_{y/x}})$, the final state of
$\NV0$ is either the $m_s=1/2$ or the $m_s=-1/2$ sublevel of the
$\qA2$ manifold. Following the same reasoning as above one can show
that optical matrix elements for these transitions are of the type
$\sqrt{2/3}\matrixel{e_{y/x}}{\vec{r}}{\phi_c}$. The resulting cross
section is identical to the cross section in the case of $m_s=\pm 1$
states. Relative transition probabilities between spin sublevels of
$\tE$ and $\qA2$ manifolds are summarized in Fig.~\ref{fig:qA2}.

\subsubsection{Photoionization from the $\sE$ state\label{sec:sE2}}

The two orbital components of the $\sE$ state can be written as
$\ket{\sE_{x}}=1/\sqrt{2}(\ket{\a\ab e_{x}\exb} - \ket{\a\ab\ey\eyb})$
and
$\ket{\sE_{y}}=1/\sqrt{2}(\ket{\a\ab \exb\ey} -
\ket{\a\ab\ex\eyb})$~\cite{Doherty2011}. Following the same reasoning
as above, one can show that after ionization the NV center transitions
to any of the four states of the $\dE$ manifold of $\NV0$ with equal
probability. The photoionization process is described by matrix
elements of the type
$1/\sqrt{2} \matrixel{e_{x/y}}{\vec{r}}{\phi_{c}}$.  Due to a
multi-determinant nature of $\sE$, the application of a simple
$\Delta$SCF procedure to calculate its energy is not
straightforward. From the theoretical standpoint, there is no
consensus regarding the position of this state above the $\tA2$ ground
state~\cite{doherty2013}. However, as discussed in Sec.~\ref{sec:sE},
energy difference of
$\left[E\left(\sE\right) - E\left(\tA2\right) \right]=0.38$~eV was
obtained in Ref.~\cite{goldman2015}.  The main focus of the current
paper is photoionization from the triplet states, so in the case of
photoionization from $\sE$ our calculations will be more
approximate. To make calculations possible we will assume that optical
matrix elements are identical to those of the photoionization from the
$\tA2$ state. In addition, we will use the same spectral function
$A(\epsilon)$ as for the ground state, ignoring the occurrence of the
Jahn-Teller effect in the $\sE$ state. The resulting cross section is
then nearly identical to of the $\tA2$ state, with the only difference
that the energies that appear in Eqs.~\eqref{sigma1} and
\eqref{sigma2} differ for the two processes.

\subsubsection {Calculation of optical matrix elements and transition energies}

The application of the Slater--Condon rule enables us to evaluate the
transition dipole moment $\vec{r}_{ij}$ for Kohn-Sham states $\psi_i$
and $\psi_j$ rather than many-electron states $\Psi_i$ and $\Psi_j$.
As the position operator is not well-defined in periodic boundary
conditions, matrix elements are calculated as
$\vec{r}_{ij} = \langle u_i|i\vec{\nabla}_{\vec{k}}|u_j\rangle$, were
$u_i$ and $u_j$ are lattice-periodic parts of wavefunctions $\psi_i$
and $\psi_j$. Eq.~\eqref{sigma1} is often alternatively formulated in
terms of momentum matrix elements~\cite{ridley}, defined as
$\vec{p}_{ij}=i m (\epsilon/\hbar) \vec{r}_{ij}$.

When replacing the many-electron formulation with the formulation
based on Kohn-Sham states, $E_{ij}$ in Eq.~\eqref{sigma1} is the
difference between Kohn--Sham eigenvalues of the defect state and the
perturbed bulk state. Since the smallest value of $E_{ij}$ does not
necessarily correspond to photoionization thresholds, obtained from
total energy calculations [$\textrm{IP}(\tA2)$ or $\textrm{IP}(\tE)$],
we apply a rigid shift so that calculated cross sections are
consistent with thresholds.  As discussed in Sec.~\ref{sec:sE2}, the
calculations for the $\sE$ state are more approximate. In this case we
use Kohn-Sham states pertaining to the $\tA2$ ground state, but the
rigid shift of energies corresponds to the ``experimental'' threshold
$\textrm{IP}(\sE)$.

\subsection{The choice of the charge state}

When replacing many-electron wavefunctions with single-particle ones,
an important issue arises regarding the charge state for which
single-particle energies and single-particle Kohn--Sham states are
calculated. On the one hand, since a negatively charged defect is the
one that is being ionized, performing calculations for a defect in the
$q=-1$ charge state could seem natural. However, while the defect
wavefunctions are represented correctly in this charge state,
conduction band wavefunctions are not. Indeed, the final state
$\psi_j$ is a conduction band state perturbed by the {\it neutral\/}
defect, and not a negative one. As a result of long-range Coulomb
interactions these perturbations are much more significant for the
negatively charged defect. On the other hand, supercell calculations
of the neutral NV center adequately capture perturbations to the
conduction bands, but these calculations do not give a totally
accurate account of the initial defect state.  The question is then:
which of the two calculations approximates the calculation based on
many-electron wavefunctions better?  Due to a localized nature of
defect states we can expect that the difference of their
single-particle Kohn--Sham wavefunctions calculated in two charge
states, $q=0$ and $q=-1$, is much smaller than the corresponding
difference between delocalized perturbed bulk states. Following the
methodology of Ref.~\cite{turiansky2020nonrad} we estimated overlap
integrals between defect levels in the case of the neutral and the
negatively charged defect.  Our result show that more than 99.6\% of
the wavefunction character is preserved when the charge state is
changed. We conclude that performing calculations in the neutral
charge state is a much more accurate approximation. This approximation
will be employed in this paper.

\subsection{Brillouin zone integration and supercell
  effects\label{sec:brillouin}}

In the supercell formulation, one obtains an appropriately normalized
cross section if one replaces the sum over $j$ in Eq.~\eqref{sigma1}
with the sum over $k$-points of the Brillouin zone of the supercell
via $\sum_j \rightarrow 1/N \sum_{n,\vec{k}}$, where $N$ is the number
of uniformly distributed $k$-points and $n$ runs over all conduction
bands states for a fixed $\vec{k}$. Matrix elements $\vec{r}_{ij}$ are
calculated between the defect state and the perturbed conduction band
state (normalized in the supercell) for the same $\vec{k}$. In
practice the sum is performed in the irreducible wedge of the
Brillouin zone.

In order to converge the cross section $\sigma_{\rm{ph}}(\epsilon)$
for a given supercell, a very dense $k$-point mesh is
required. Increasing the mesh in self-consistent calculations of
supercells becomes computationally very expensive even at the PBE
level. Charge density converges much faster as the $k$-point mesh is
increased. Thus, we performed self-consistent calculations using the
$6 \times 6 \times 6$ Monkhorst-Pack $k$-point mesh for the charge
density. Photoionization cross sections have been calculated by
performing non-self-consistent calculations using much denser
$14 \times 14 \times 14$ meshes.  In this way one obtains
photoionization cross sections of a periodically repeated array of NV
centers (albeit correctly normalized per one absorber).

\begin{figure}
\begin{center}
\includegraphics[width=0.99\linewidth]{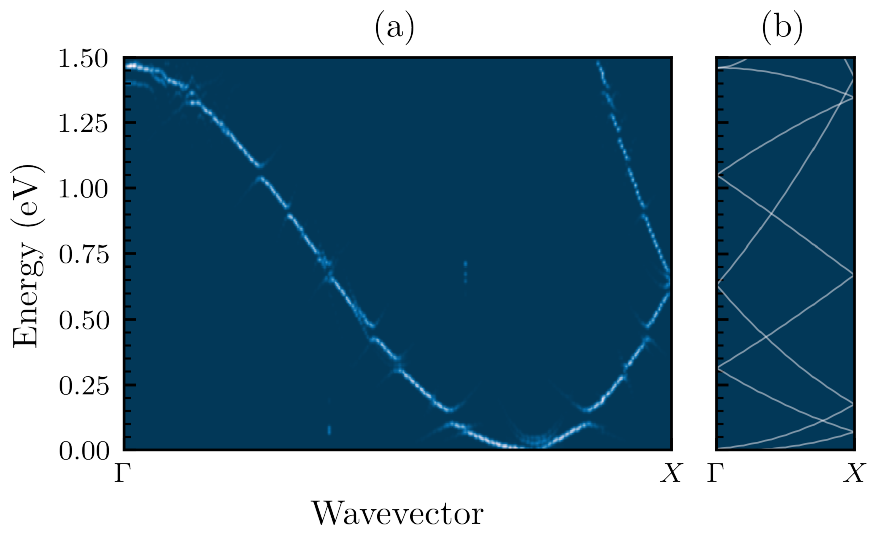}
\caption{(a) Unfolded band structure of conduction band states
  perturbed by the NV center along the $\Gamma\mbox{--}X$ path. The
  color indicates a relative spectral weight (dark blue is zero), see
  Ref.~\cite{popescu2012} for more details. (b) The band structure of
  bulk diamond, folded to the first Brillouin zone of the
  $4\times4\times4$ supercell along the $\Gamma\mbox{--}X$ path of the
  supercell.\label{fig:k-depend}}
\end{center}
\end{figure}

Artificial periodicity of the supercell approach gives rise to two
undesirable effects: (i) defect--defect interaction and (ii)~spurious
perturbation of conduction band states. Aspect (i)~affects defect
wavefunctions. To check the convergence of these wavefunctions as a
function of the supercell size one can, for example, calculate the
optical matrix element $\vec{r}_{ij}$ [Eqs.~\eqref{stimulated}
and~\eqref{intra}] for the transition between $a_1$ and $e$ levels of
the NV center.  The comparison of $4\times 4 \times 4$ and
$5 \times 5 \times 5$ supercells shows that matrix elements calculated
in these two supercells differ by less than 3\%.

Effect (ii), however, is more subtle. Periodically repeated NV centers
form a superlattice and one could expect the formation of sub-bands
and the opening of ``mini-gaps'' in the same way it occurs in
traditional semiconductor superlattices. This is indeed what we
observe.  In Fig.~\ref{fig:k-depend}(a) we show the band structure of
the $4 \times 4 \times 4$ supercell unfolded~\cite{popescu2012} onto
the Brillouin zone of a primitive diamond cell. For illustration
purposes we choose the band structure of a neutral NV in the $\qA2$
state. In Fig.~\ref{fig:k-depend}(a) one can clearly identify
discontinuities in the band structure. To understand why these
discontinuities form at specific energies and $k$ vectors, in
Fig.~\ref{fig:k-depend}(b) we show the band structure of bulk diamond
{\it folded\/} onto Brillouin zone of the $4 \times 4 \times 4$
supercell.  Such folding introduces degeneracies at the band crossing
points and Brillouin zone boundaries. When perturbations, such as the
potential of periodically repeated NV centers, are present, these
degeneracies are removed, explaining the formation of ``mini-gaps'' in
Fig.~\ref{fig:k-depend}(a). Apart from the density of states (DOS), we
find that the values optical matrix element $\vec{r}_{ij}$ are also
affected by artificial periodicity. We observe jumps of $r^2_{ij}$
across the ``mini gaps''. These jumps can be explained using the
textbook picture of the behavior of electronic wavefunctions close to
the band gap in pristine solids via the formation of standing
electronic waves (see, e.g., Fig.~3 in Chapter~7 of
Ref.~\cite{kittel}). The wave-function on one edge of the ``mini-gap''
has a vanishing weight on the NV center and $\vec{r}_{ij}$ pertaining
to this state tends to zero. The wavefunction on the other edge has
maximum weight on the NV center, and $\vec{r}_{ij}$ pertaining to that
state attains a finite value.  We conclude that artificial periodicity
affects both the energies of conduction band states and the values of
optical matrix elements. This is the origin for our observed slow
convergence of calculated cross sections as a function of the
supercell (not shown), even when the Brillouin zone integration itself
is already converged.

In this paper we use the following {\it ad hoc\/} solution to this
problem.
(i)~Each perturbed conduction band state of the defect supercell is
unfolded to the Brillouin zone of the primitive cell using the
methodology of Ref.~\onlinecite{popescu2012}.  Each $k$-point of the
Brillouin zone of the defect supercell unfolds onto several $k$-points
of the Brillouin zone of the primitive cell.
(ii)~We take the $k$-vector with the highest spectral weight and find
the bulk state with the same $k$ which is closest in energy (typical
differences $<0.08$~eV).  This is the energy that we use in
Eq.~\eqref{sigma1}.  In this way we get rid of the discontinuities of
the conduction band state energies.  The procedure also yields the
band index~$n$.  In the case of degeneracy this $k$-point is assigned
to multiple $n$'s.
(iii)~For a given perturbed conduction band state the value of the
optical matrix element $r^2_{ij}$ is averaged in the Brillouin zone of
the primitive cell. The averaging is performed for each $n$
separately, taking the mean of the points situated closer than
$\Delta r_{k} = 0.57$~nm$^{-1}$ in the reciprocal lattice.  This
``smears'' the jumps of the optical matrix elements across ``mini
gaps''.  The overall procedure results in smooth calculated values of
$\sigma_{\rm{ph}}(\epsilon)$ as a function of $\epsilon$. The effect
of such smoothing is illustrated in Sec.~II of the Supplemental
Material~\cite{supp}.

\subsection{Local-field effects\label{sec:local-field}}

In the expressions for photoionization cross
sections~\eqref{sigma1},~\eqref{stimulated}, and~\eqref{intra} we have
omitted so-called local-field effects \cite{stoneham,ridley}.  These
effects appear due to the scattering of light on the defect, which can
result in the electric field on the defect site being different from
that in the bulk.  Historically, these effects have been included
phenomenologically by multiplying the cross sections by an enhancement
factor ${(\mathcal{E}_{\rm{eff}}/\mathcal{E}_{0})}^2$.  Here
$\mathcal{E}_{\rm{eff}}$ is the electric field on the defect site,
while $\mathcal{E}_{0}$ is the electric field in the bulk.  Classical
considerations lead to a variety of models~\cite{ridley,stoneham},
from which the so-called Onsager model typically performs best, even
though it slightly overestimates the enhancement factor (see
Table~10.3 in Ref.~\cite{stoneham}). In~the Onsager model the ratio
between fields is given by:
\begin{equation}
  \frac{\mathcal{E}_{\rm{eff}}} {\mathcal{E}_{0}}
  =
  \frac{3 \varepsilon_{\infty}}{2\varepsilon_{\infty} + 1},
\end{equation}
where $\varepsilon_\infty$ is the dielectric constant of
diamond. Using $\varepsilon_\infty=5.7$, we obtain
$(\mathcal{E}_{\rm{eff}}/\mathcal{E}_{0})=1.38$.

An alternative method to estimate local-field effects is
empirical. Radiative emission rate for the ${\tE} \to {\tA2}$
transition is given via:
\begin{equation}
  \Gamma_{\rm{rad}} =\frac{1}{\tau_{\rm{rad}}}
  =
  {\left(
    \frac{\mathcal{E}_{\rm{eff}}} {\mathcal{E}_{0}}
  \right)}^2
\frac{n_D E_{\rm{ZPL}}^{3} r_{ij}^2} {3\pi\varepsilon_0 c^3 \hbar^4}.
\label{eq:spont}
\end{equation}
Here $\varepsilon_{0}$ is vacuum permittivity, and $r_{ij}$ is the
transition dipole moment for the transition ${\tE}\to{\tA2}$.
Comparing the value calculated without local-field effects with the
experimental result can provide an estimate for
$(\mathcal{E}_{\rm{eff}}/\mathcal{E}_{0})$. Employing the PBE
functional, our calculated radiative lifetime without local field
effects is $\tau_{\rm{th}}=12.2$~ns (using the experimental ZPL
energy), in excellent accord with the experimental value $\tau=12$~ns
\cite{doherty2013}. This yields
$(\mathcal{E}_{\rm{eff}}/\mathcal{E}_{0}) \approx 1.01 $. We conclude
that the Onsager model over-estimates the value of
$(\mathcal{E}_{\rm{eff}}/\mathcal{E}_{0})$, in accord with the results
for F-centers in alkali halides~\cite{stoneham}. Note, however, that
the theoretical value (and therefore possible differences with
experiment) are affected not only by the inclusion/exclusion of local
fields, but also by other approximations that we employed (density
functionals, calculations of matrix elements using Kohn--Sham states,
etc.). Regardless, we estimate that in the case of NV centers
$(\mathcal{E}_{\rm{eff}}/\mathcal{E}_{0})$ is in the range
$(1\mbox{--}1.4)$, and very likely close to $1$. In the remainder of
this paper we will therefore set the enhancement factor
${(\mathcal{E}_{\rm{eff}}/\mathcal{E}_{0})}^2$ to 1.


\section{Results\label{sec:results}}

\subsection{Excitation energies and photoionization thresholds}

For the photoionization threshold from the ground state $\tA2$ we
obtain the value $\textrm{IP}(\tA2)= 2.67$~eV. This is close to the
previously published {\it ab-initio\/} result of
2.64~eV~\cite{deak2014}.  We attribute a small difference of 0.03~eV
to a slightly larger kinetic energy cutoff used in the current work,
and, probably more importantly, a different finite-size correction
scheme (cf.~Ref.~\cite{freysoldt2009}). Both of the calculated
thresholds are in a very good agreement with the experimental one of
${\sim}2.6$~eV, determined in Ref.~\cite{aslam2013}. This establishes
an error bar of about $0.1$~eV for the agreement of {\it ab-initio\/}
calculations with experimental data regarding photoionization
thresholds.

The ZPL energy of the intra-defect transition ${\tA2}\rightarrow{\tE}$
is found to be $E(\tE)- E(\tA2)=1.996$~eV. This is again in agreement
with previous calculations~\cite{gali2009} and the experimental value
of $1.945$~eV, exhibiting an accuracy better than $0.1$ eV. Finally,
for the energy difference $[E(\qA2)-E(\dE)]$ in the case of $\NV0$ we
obtain the value of $0.48$~eV. As discussed in
Sec.~\ref{sec:threshold}, the experimental energy difference is not
available. Previous calculations based on the diagonalization of the
Hubbard Hamiltonian (albeit with a rather small
basis)~\cite{ranjbar2011} yielded a value of 0.68 eV for the vertical
transition (i.e., keeping the atoms fixed in the geometry of the
$\dE$) state. Including our calculated relaxation energy of 0.12~eV we
obtain a corrected value of 0.56~eV, in good agreement with our
result.  Eventually, using Eq.~\eqref{IP2}, we obtain the
photoionization threshold from the $\tE$ state $\textrm{IP}(\tE)=1.15$
eV. Calculated and experimental thresholds for photoionization from
$\tA2$, $\tE$, and $\sE$ states are summarized in Table~\ref{tab:IP}.
Photoionization threshold from the $\sE$ state has not been measured
directly, but is deduced from the analysis in Ref.~\cite{goldman2015}.

\begin{table}
\begin{tabular}{@{}c c c c}
\hline
            &  $ \textrm{IP}(\tA2) $  & $ \textrm{IP}(\tE) $ & $ \textrm{IP}(\sE) $ \\
\hline
theory      &  2.67              &  1.15 & -- \\
experiment  &  2.6               &  --  & {\it 2.2} \\
\hline
\end{tabular}
\caption{Thresholds for photoionization from the $\tA2$, 
$\tE$, and $\sE$ states of $\NVm$ (in eV). The experimental result for $\textrm{IP}(\tA2)$
  from Ref.~\cite{aslam2013}. The value of $\textrm{IP}(\sE)$ has not been measured directly, but deduced
  from Ref.~\cite{goldman2015} (shown in italic). The assumed experimental error bar of the latter two thresholds
  is 0.1 eV. 
  \label{tab:IP}}
\end{table}

\begin{figure*}
\includegraphics[width=0.99\linewidth]{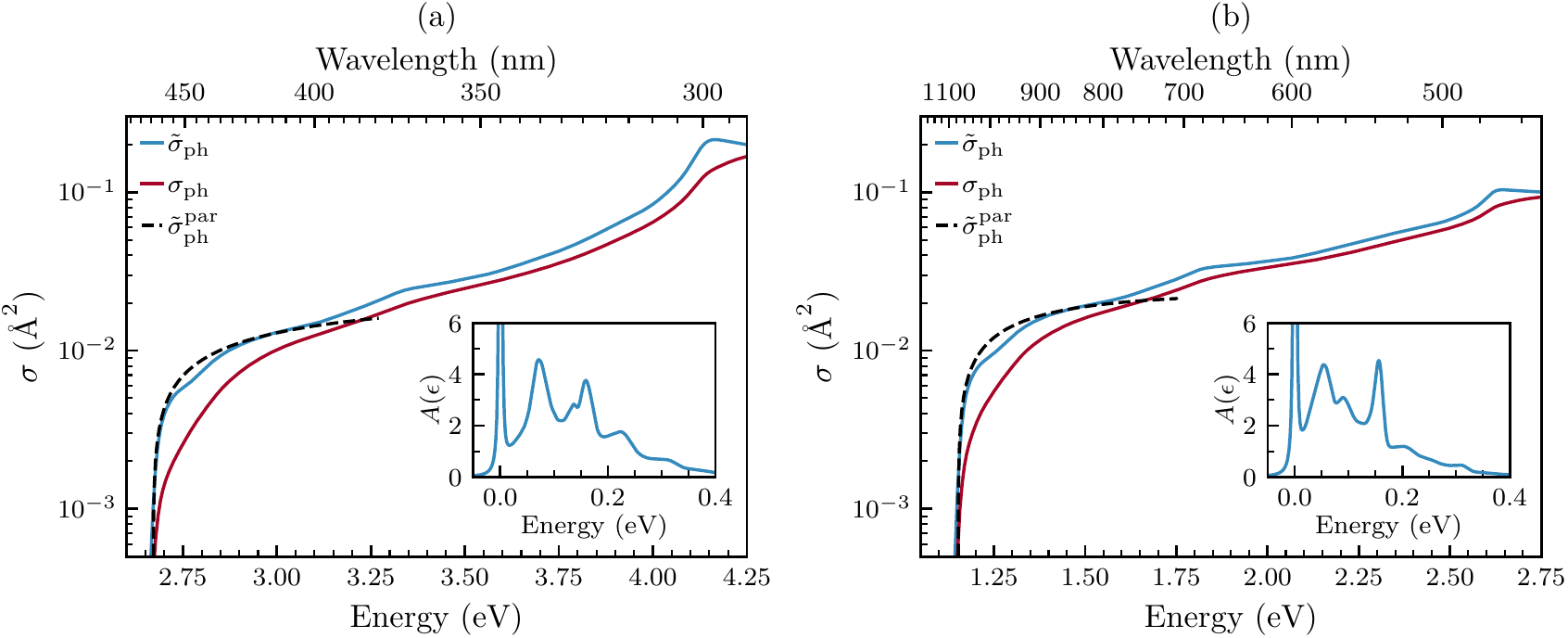}
\caption{Photoionization cross sections from (a) the $\tA2$ and (b)
  the $\tE$ state of $\NVm$. Blue lines: cross sections
  $\tilde{\sigma}_{\rm{ph}}(\epsilon)$ without vibrational broadening
  [Eq.~\eqref{sigma1}]; red lines: actual cross sections
  $\sigma_{\rm{ph}}(\epsilon)$ [Eq.~\eqref{sigma2}]; dashed lines show
  $\tilde{\sigma}_{\rm{ph}}$ calculated using a constant momentum
  matrix element in Eq.~\eqref{sigma1} and DOS corresponding to a
  parabolic band (see text).  Insets show the spectral function of
  electron--phonon coupling $A(\epsilon)$.\label{fig:cross_ground}}
\end{figure*}

\subsection{Cross sections\label{sec:res_cross}}

Before we present the results for NV centers, let us first briefly
review some aspects regarding the existing knowledge of
photoionization cross sections of deep defects in
solids~\cite{ridley}. The overall shape of the function
$\tilde{\sigma}_{\rm{ph}}(\epsilon)$ depends on the specifics of the
defect wavefunction and of the bulk conduction band structure. Over
the years many analytical and semi-analytical models of
photoionization of deep defects have been
developed~\cite{ridley}. Most of these models use the formulation
based on the momentum matrix element
$\vec{p}_{ij}=i m (\epsilon/\hbar) \vec{r}_{ij}$, which we will also
follow in this Section.

For the conduction band with a parabolic dispersion close to the CBM,
photoionization threshold corresponds to the excitation to the CBM
with the electronic density of states
$D(E)\sim {\left(E-E_{\rm{CBM}}\right )}^{1/2}$.  In this situation
there are two limit cases regarding the dependence of the momentum
matrix element $\vec{p}_{ij}$ on $\Delta \vec{k}$, where
$\Delta \vec{k}$ is the quasi-momentum measured with respect to the
value at the CBM.\@ One limit corresponds to a system where the
character of the defect wavefunction is essentially the same as the
character of bulk states near the CBM.\@ In~this case one
obtains~\cite{ridley} $\vec{p}_{ij}\sim\Delta \vec{k}$, which yields
the cross section (without electron-phonon coupling) close to the
threshold
$\tilde{\sigma}_{\rm{ph}}(\epsilon)\sim$${(\epsilon-\epsilon_{\rm{th}})}^{3/2}$.
Here
$\epsilon_{\rm{th}}$ is the threshold for photoionization, i.e.,
$\textrm{IP}(\tA2)$, $\textrm{IP}(\tE)$, or
$\textrm{IP}(\sE)$ from Table~\ref{tab:IP}.  One could say that the
transition from the defect state to the CBM is dipole-forbidden. Such
a scenario is described, e.g., by the widely-used Lucovsky
model~\cite{lucovsky1965}.  Another limit describes a dipole-allowed
transition to the CBM.\@ This happens, for example, when the defect
state has $p$ character, while the conduction band states have
$s$ character, or vice versa. In this case, to a very good
approximation, $\vec{p}_{ij}$ is constant for small $\Delta
\vec{k}$~\cite{ridley}, and one obtains $\tilde{\sigma}_{\rm{ph}}(
\epsilon)\sim$${(\epsilon-\epsilon_{\rm{th}})}^{1/2}$ close to the
absorption edge.

\begin{figure*}
\begin{center}
\includegraphics[width=0.9\linewidth]{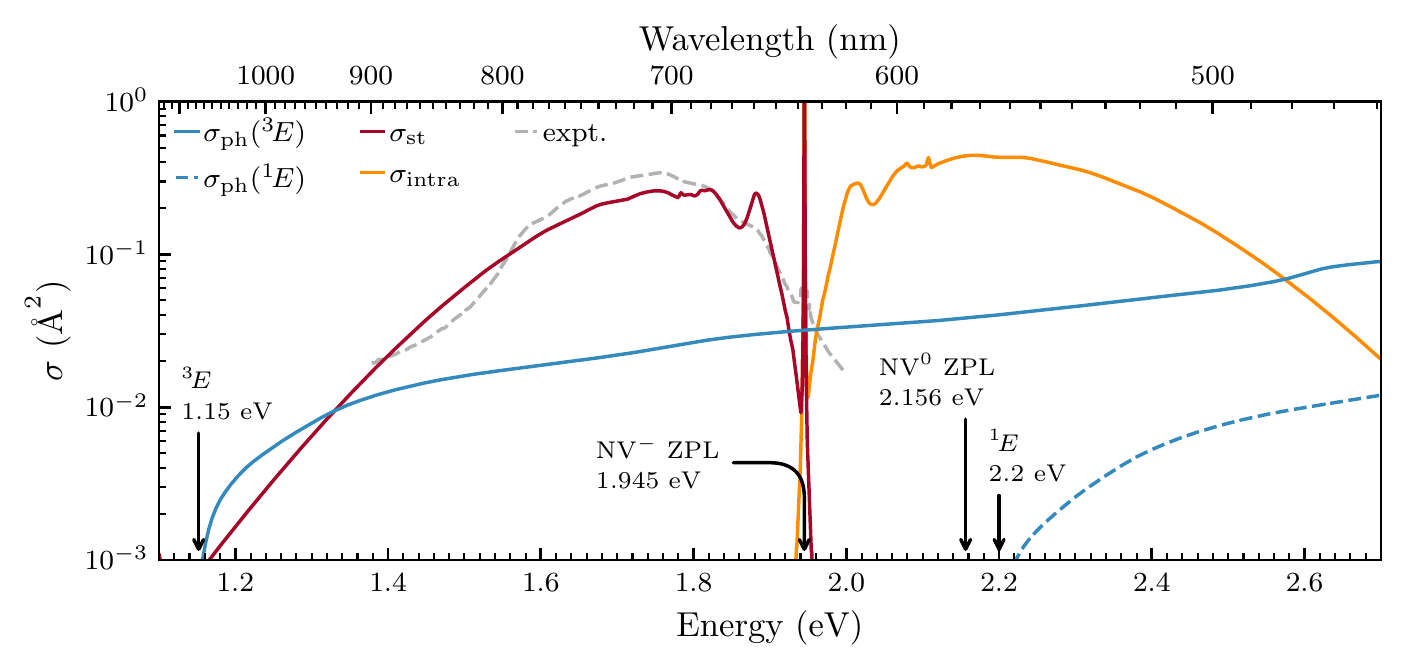}
\caption{Calculated cross section as a function of photon
  energy. Solid blue: photoionization from the excited state $\tE$,
  $\sigma_{\rm{ph}}$; dark red: stimulated emission,
  $\sigma_{\rm{st}}$; orange: intra-defect absorption,
  $\sigma_{\rm{intra}}$; dashed blue: photoionization from the singlet
  state $\sE$. Dashed gray line shows cross section for stimulated
  emission from Ref.~\cite{jeske2020}.  Photoionization threshold from
  $\tE$ and $\sE$ are indicated (estimated error bar 0.1 eV), together
  with the experimental values of the ZPL energy for $\NVm$ and
  $\NV0$.\label{fig:cross_exc}}
\end{center}
\end{figure*}

The photoionization cross section $\tilde{\sigma}_{\rm{ph}}(\epsilon)$
for the ground state $\tA2$ is shown in Fig.~\ref{fig:cross_ground}(a)
(blue solid line). $\delta$~functions in Eq.~\eqref{sigma1} have been
replaced by Gaussians with width $\sigma=30$~meV. We find that near
the threshold
$\tilde{\sigma}_{\rm{ph}}(\epsilon)$$\sim$${(\epsilon-\epsilon_{\rm{th}})}^{1/2}$,
indicating that the transition to the band edge is dipole-allowed and
the momentum matrix element near the CBM attains a constant value (for
more details, see Sec.~III of the Supplemental Material~\cite{supp}).
$\tilde{\sigma}_{\rm{ph}}(\epsilon)$ pertaining to this constant value
of the momentum matrix element and DOS corresponding to a parabolic
band is also shown in Fig.~\ref{fig:cross_ground}(a) (dashed line).
The parabolic dispersion is characterized by effective electron masses
$m_{\parallel}$ and $m_{\perp}$. In our calculations we used our
obtained theoretical values $m_{\parallel}=1.66 m_e$ and
$m_{\perp}=0.32 m_e$ that are in a good agreement with experimental
ones~\cite{nava1980}. At larger photon energies
$\tilde{\sigma}_{\rm{ph}}(\epsilon)$~starts to deviate from the
$\sim$${(\epsilon-\epsilon_{\rm{th}})}^{1/2}$ behavior because the DOS
of conduction band states departs from that of the parabolic band and
momentum matrix elements start to deviate from a value at the
threshold \cite{supp}.  Lastly and most importantly,
Fig.~\ref{fig:cross_ground}(a) shows the actual photoionization cross
section
$\sigma_{\rm{ph}}(\epsilon)$ (dark red line) that includes the effects
of vibrational broadening; the inset depicts the spectral function of
electron--phonon coupling $A(\hbar
\omega)$~\cite{razinkovas2020,supp}.  Vibrational broadening shifts
the weight of the cross section to higher energies and
$\sigma_{\rm{ph}}(\epsilon)$ no longer exhibits the
$\sim$$(\epsilon-\epsilon_{\rm{th}})^{1/2}$ behavior close to the
absorption edge.  In~passing, we note that the results regarding a
constant value of $p_{ij}$ around the CBM confirm the assumptions and
the value of the momentum matrix element used in our recent
calculations on NV centers in diamond nanowires~\cite{oberg2019}.

Calculated cross sections for the photoionization from the $\tE$ state
are shown in Fig.~\ref{fig:cross_ground}(b). The behavior of
$\tilde{\sigma}_{\rm{ph}}(\epsilon)$ as a function of photon energy
$\epsilon$ (blue line) can be explained similarly as for the ground
state. In short: (i)~close to the threshold
$\tilde{\sigma}_{\rm{ph}}(\epsilon)$$\sim$$(\epsilon-\epsilon_{\rm{th}})^{1/2}$
(dashed line); (ii)~at larger photon energies
$\tilde{\sigma}_{\rm{ph}}(\epsilon)$~starts to deviate from this
functional form because the electronic DOS departs from that of the
parabolic band and momentum matrix elements can no longer be assumed
constant; (iii)~actual photoionization cross section
$\sigma_{\rm{ph}}(\epsilon)$, which includes the vibrational
broadening, exhibits a blue-shift with respect to
$\tilde{\sigma}_{\rm{ph}}(\epsilon)$ (dark red line).

The main results of the current paper are presented in
Fig.~\ref{fig:cross_exc}. Therein, the photoionization cross section
from the excited state $\tE$ (solid blue line) is shown together with
the calculated cross sections for stimulation emission
$\sigma_{\rm{st}}(\epsilon)$ from the excited state $\tE$
[Eq.~\eqref{stimulated}, dark red line], absorption from the ground
state ${\tA2}$ [Eq.~\eqref{intra}, orange line], and photoionization
cross section from the $\sE$ state (dashed blue line).
Photoionization cross section from the $\tA2$ state is not shown.  The
result for $\sigma_{\rm{st}}(\epsilon)$ is in good agreement with the
one presented by Jeske~{\it et~al.\/}~in Ref.~\cite{jeske2020} (dashed
gray line).  In that paper the value of the cross section was deduced
from the expression identical to Eq.~\eqref{stimulated}, but using the
``experimental'' value of the optical matrix element and the
experimental spectral function $A(\hbar \omega)$. The optical matrix
element was calculated from the experimental value of the spontaneous
emission rate as in Eq.~\eqref{eq:spont}.  The consequences of our
findings for the physics and technology of NV centers are discussed in
the next Section.


\section{Discussion\label{sec:discussion}}

\subsection{Existing understanding of the photoionization of NV
  centers}

DFT calculations of absorption cross sections from the ground state
$\tA2$ have been previously reported in Ref.~\cite{bourgeois2017}.
Our work differs from the results of these authors in the following
aspects: (i) In our work we present absolute values for
$\sigma_{\rm{ph}}(\epsilon)$, while in Ref.~\cite{bourgeois2017} the
cross section has been determined in arbitrary units. (ii) The issues
regarding the Brillouin zone integration and supercell size
convergence (see Sec.~\ref{sec:brillouin}) had not been fully dealt
with in Ref.~\cite{bourgeois2017}, which yielded spurious oscillations
of the cross section (Fig.~3 of Ref.~\cite{bourgeois2017}).  (iii) The
coupling to phonons was included in our calculations via the spectral
function $A(\hbar\omega)$, while this coupling was omitted from
Ref.~\cite{bourgeois2017}.

The mechanism of the photoionization process from the excited state
$\tE$ has been first analyzed in Ref.~\cite{siyushev2013}. These
authors suggested that after the photoionization of $\NVm$ an Auger
process takes place, whereby an electron from the conduction band
transitions to the $a_1$ level, while an electron from the $e$ level
is excited to the conduction band. It is our conviction that Auger
capture rates have been significantly over-estimated in the
calculations of Ref.~\cite{siyushev2013}, as they were obtained
assuming exceedingly large effective electron densities
$\sim$$3\times10^{20}$
cm$^{-3}$. In our opinion, no Auger process takes place. Assuming a
realistic electron velocity of
$10^{7}$~cm/s after photoionization, one can estimate that within the
first picosecond the electron moves away from the NV center by a
distance of 100~nm, making any non-radiative capture process very
unlikely. Instead, as we propose in the current work, the NV center
directly transitions to the $\qA2$ electronic state of
$\NV0$, resulting in the expression for the threshold Eq.~\eqref{IP2}.

To the best of our knowledge, the details of the photoionization from
the $\sE$ states have not been addressed previously.

\subsection{$\qA2$ as a state of $\NV0$ directly after
  photoionization\label{sec:qA2}}

The fact that after photoionization from the $\tE$ state NV centers
transitions into the metastable $\qA2$ state of $\NV0$ has important
consequences for charge dynamics of NV centers. In particular, this
can explain ESR experiments of Ref.~\cite{felton2008}, where a signal
related to the $\qA2$ state was clearly observed. The results of that
work allowed the authors to conclude that the existence of a strong
ESR signal was due to spin polarization in the $\qA2$ manifold. One
possible explanation of spin polarization is that the inter-system
crossing from the $\dA2$ excited state of the $\NV0$ center might lead
to spin polarization the same way as it occurs for the negative charge
state.  However, we propose a different scenario which
straightforwardly follows from our results.

The spin physics of the photoionization from the $\tE$ state was
discussed in Sec.~\ref{sec:tEa}.  Fig.~\ref{fig:qA2} summarizes
transitions from different sublevels of the $\tE$ manifold of $\NVm$
to spin sublevels of the $\qA2$ manifold of $\NV0$. Numbers indicate
relative transition probabilities from a given state. In particular,
if the defect is initially in the $m_s=+1$ ($-1$) spin sublevel, the
probability of the transition to the $m_s=+3/2$ ($-3/2$) sublevel is
$3/4$, while that to the $m_s=+1/2$ ($-1/2$) sublevel is $1/4$. If the
initial spin state is $m_s=0$, after the ionization $\NV0$ is in any
of the $m_s=\pm 1/2$ spin states of the $\qA2$ manifold with equal
probability. Importantly, one can deduce from Fig.~\ref{fig:qA2} that
if $\NVm$ is initially spin-unpolarized (occupation of different spin
sublevels is equal), then there is no spin polarization of the $\qA2$
state after photoionization.

In Ref.~\cite{felton2008} the ESR signal pertaining to the $\qA2$
state was only observed for laser wavelengths above the ZPL of the
neutral NV center, 2.156~eV~\cite{felton2008}.  For such illumination
the NV center is constantly switching between the negative and the
neutral state~\cite{beha2012,aslam2013}.  When the NV center is in the
negative charge state, intrinsic processes within the electronic
states of $\NVm$ lead to a preferential population of the $m_s = 0$
spin sublevel in the $\tA2$ and the $\tE$ spin triplets
\cite{doherty2013}. Thus, photoionization mostly occurs from the
$m_s=0$ sublevel of the $\tE$ state. As per transition probabilities
shown in Fig.~\ref{fig:qA2}, $m_s=\pm 1/2$ spin sublevels of the
$\qA2$ are preferentially occupied after the photoionization.
$m_s = \pm 1/2$ and $m_s = \pm 3/2$ spin sublevels are separated by a
zero-field splitting $D (\qA2) = 1.69$~GHz~\cite{felton2008}, and thus
the population of $m_s = \pm 1/2$ sublevels lays ground to a strong
ESR signal. In fact, we take the experimental results of
Ref.~\cite{felton2008} as an indirect confirmation of our proposal
regarding the involvement of the $\qA2$ state in the photoionization
from the $\tE$ state. Simply put, spin polarization of the $\qA2$
state of $\NV0$ found in Ref.~\cite{felton2008} directly stems from
spin polarization of $\NVm$.

\subsection{Spin-to-charge conversion under dual-beam excitation}

Turning now to our calculated cross sections and their relevance to
the photophysics of NV centers, we emphasize that the body of
experimental work on charge-state dynamics at NV centers is large. To
consistently interpret all of that work the knowledge of
photoionization of $\NVm$ is often insufficient and the knowledge of
similar processes for $\NV0$ is needed. This is especially true for
steady-state experiments, where, depending on the wavelength of
laser(s), the NV center can constantly switch between the two charge
states.  As the study of $\NV0$ is beyond the scope of the current
work, we will focus only on a few selected experiments that we are now
able to explain using our data.  It is important to stress at the
outset that the quality of samples in these studies is crucial.  We
discuss only experiments done on high-quality bulk samples and will
not mention numerous experiments performed on nanodiamonds. The
existence of surfaces and possibly various surface defects makes the
charge-state dynamics of NV centers in nanodiamonds highly complex and
not always reproducible.

One important set of such experiments is spin-to-charge conversion
upon dual-beam excitation at cryogenic temperatures
\cite{Zhang2021,Irber2021}.  In these protocols one narrow laser is
used to resonantly excite $\NVm$ in a pre-selected spin state, e.g.,
$m_s=0$. Another laser pulse is then used to photoionize the defect
from the $\tE$ state. In order not to disturb other spin states, the
photoionization is performed with sub-ZPL illumination. In the work of
Irber {\it et al.}  efficient photoionization was obtained using a
visible laser emitting at 642 nm (1.93~eV)~\cite{Irber2021}.
At~variance, Zhang {\it et al.} used a NIR laser emitting at 1064 nm
(1.17~eV)~\cite{Zhang2021}; the overall scheme reached high spin
read-out fidelities, implying a rather effective photoionization.
These results naturally prompt the question: are these two energies,
1.17 eV and 1.93 eV, special?

\begin{figure}
\begin{center}
\includegraphics[width=0.99\linewidth]{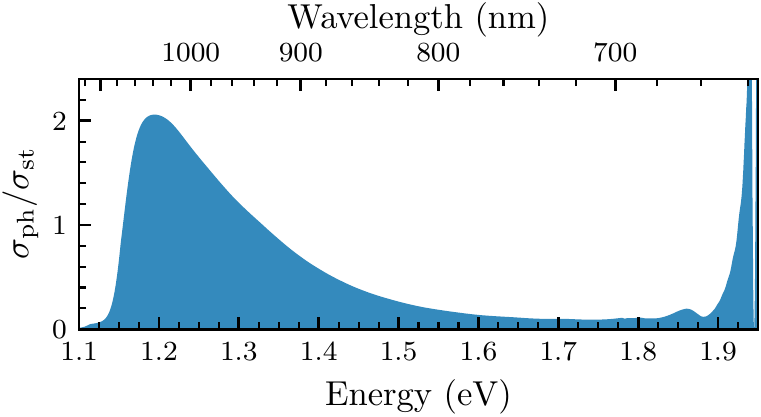}
\caption{The ratio of the photoionization cross section and the cross
  section for stimulated emission,
  $\sigma_{\textrm{ph}}/\sigma_{\textrm{st}}$, as a function of photon
  energy.\label{fig:ratio}}
\end{center}
\vspace{-1em}
\end{figure}
	
We can now answer this question using the results of our
calculations. In Fig.~\ref{fig:ratio} we plot the ratio of the
photoionization cross section and that for stimulated emission as a
function of photon energy, both from Fig.~\ref{fig:cross_exc}. One can
clearly identify two regions where photoionization of $\NVm$ with
sub-ZPL photons is most efficient
($\sigma_{\textrm{ph}}/\sigma_{\textrm{st}}>1$): (i) just above the
threshold energy of 1.15~eV, but below $\sim$1.3 eV; (ii) just below
the ZPL of 1.945~eV.  These are exactly the two energy ranges for
which the photoionization was successful in the experiments of
Refs.~\cite{Zhang2021,Irber2021}.  We note, however, that the exact
value of $\sigma_{\textrm{ph}}/\sigma_{\textrm{st}}$ in these two
energy windows depends quite sensitively on possible errors in our
calculated value of $\textrm{IP}(\tE)$.  Despite this, we conclude
that our calculations provide a consistent explanation of the results
of Refs.~\cite{Zhang2021} and~\cite{Irber2021}.  Our results point to
photon energies where spin-to-charge conversion with sub-ZPL
illumination is most efficient. We note that in this work we
considered photoionization of an ensemble of randomly oriented NV
centers (or, alternatively, photoionization with unpolarized light).
In~the case of single centers (effectively, excitation and
photoionization with polarized light) the ratio
$\sigma_{\textrm{ph}}/\sigma_{\textrm{st}}$ depends strongly on
polarization. Photoionization of single NVs will be discussed
elsewhere.

Finally, we are now in the position to comment on the results of
Ref.~\cite{jeske2017}, already mentioned in the Introduction,
Sec.~\ref{sec:intro}. Studying the luminescence of $\NV0$ upon
illumination with a cw green laser vs.~a cw green laser and a pulsed
red laser the authors observed that photoionization starts to dominate
stimulated emission at below the wavelength of the red laser $\sim$640
nm (1.938~eV).  The photoionization threshold from the $\tE$ state was
estimated to be $\sim$1.88 eV. The first result is in full agreement
with our study.  As we show in Fig.~\ref{fig:ratio}, photoionization
indeed dominates stimulated emission just below the ZPL of
1.945~eV. However, our results show that the threshold for
photoionization is 1.15 eV, and not 1.88 eV.  The value of 1.88 eV
might correspond to the dip in the ratio
$\sigma_{\textrm{ph}}/\sigma_{\textrm{st}}$ occurring at about 1.88 eV
(Fig.~\ref{fig:ratio}). This dip is due increased stimulated emission
at this energy (Fig.~\ref{fig:cross_exc}), which in its turn simply
reflects the first phonon side-peak in luminescence.


\section{Conclusions\label{sec:conclusions}}

In this paper we presented {\it ab-initio} calculations of
photoionization thresholds and cross sections
$\sigma_{\rm{ph}}(\epsilon)$ for the negatively charged
nitrogen--vacancy center in diamond. From the point of view of
computational materials science, our work presented a new methodology
to calculate photoionization cross sections. We employed an
integration on a dense $k$-point mesh, together with band unfolding
and interpolation, to obtain smooth functions
$\sigma_{\rm{ph}}(\epsilon)$ over the entire energy range.  To the
best of our knowledge, this is the first calculation of absolute
photoionization cross sections for point defects using modern
electronic structure methods. The methodology is directly applicable
to other point defects, including quantum
defects~\cite{bassett2019}. From the point of view of NV physics, we
showed that right after the photoionization from the $\tE$ state the
$\NVm$ transitions into the $\qA2$ state of $\NV0$. This explains spin
polarization observed in electron spin resonance experiments of the
$\qA2$ state. We determine that the photoionization threshold from the
$\tE$ state is 1.15~eV. Together with calculated cross sections, this
explains recent experiments on spin-to-charge conversion based on
dual-beam excitation~\cite{Zhang2021,Irber2021}. Our work provides
important new knowledge about charge-state dynamics of NV centers that
was hitherto been missing.

\vspace{0.3cm}
\section*{Acknowledgements} 
\vspace{-0.2cm}
We acknowledge A. Gali for discussions.
This work has been funded from the European Union's Horizon 2020
research and innovation programme under grant agreement No.~820394
(project A{\sc steriqs}).  MM acknowledges support by the National
Science Centre, Poland (Contract 2019/03/X/ST3/01751) within the M{\sc
  iniatura} 3 Program.  MWD acknowledges support from the Australian
Research Council (DE170100169).  Computational resources were provided
by the High Performance Computing center ``HPC Saul\.etekis'' in the
Faculty of Physics, Vilnius University and the Interdisciplinary
Center for Mathematical and Computational Modelling (ICM), University
of Warsaw (grant No.~GB81-6).




\bibliography{references} 

\end{document}


\title[NV photoionization]{Photoionization of diamond NV centers:
  theory and {\it ab initio\/} calculations. Supplemental material}

\author{Lukas Razinkovas}

\author{Marek Maciaszek} 

\author{Friedemann Reinhard}

\author{Marcus W. Doherty}

\author{Audrius Alkauskas}

\noaffiliation
\maketitle	

\section{Spectral functions of electron--phonon coupling $A(\epsilon)$}

\begin{figure}[h]
\begin{center}
\includegraphics[width=0.95\linewidth]{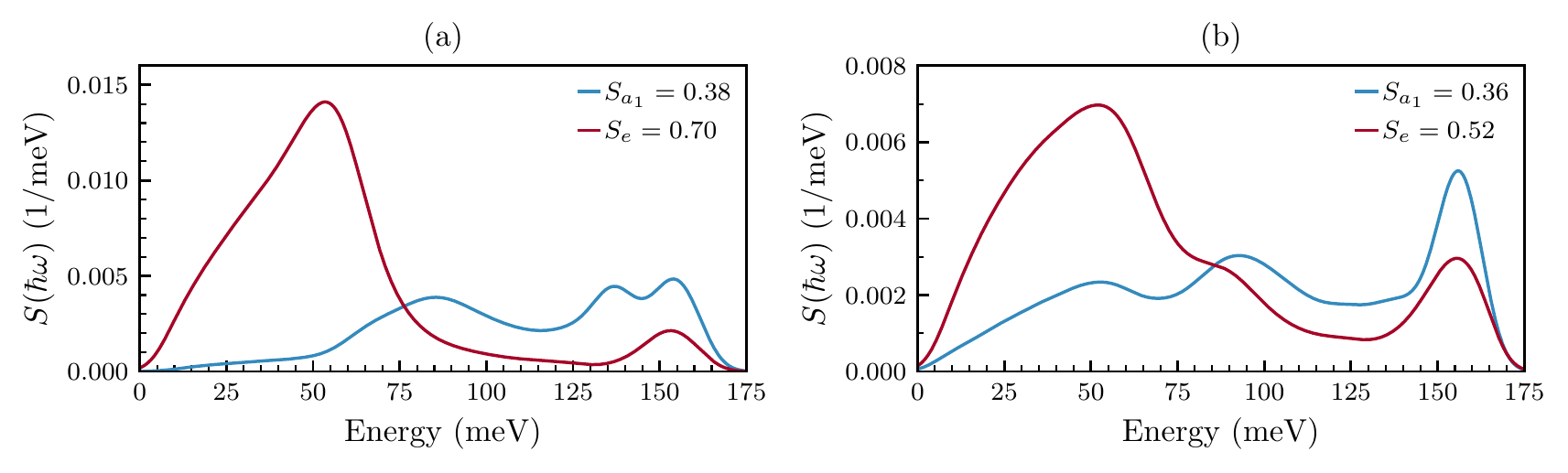}
\caption{Spectral densities of electron--phonon coupling $S(\hbar\omega)$ for $a_1$ (blue) and $e$ (dark red) vibrations: (a) $\tA2 \to \dE$ transition
(photoionization from $\tA2$);
(b) $\tE \to \qA2$ transitions (photoionization from $\tE$). \label{fig:hr_spectrum}}
\end{center}
\end{figure}

\begin{figure}
\begin{center}
\includegraphics[width=0.95\linewidth]{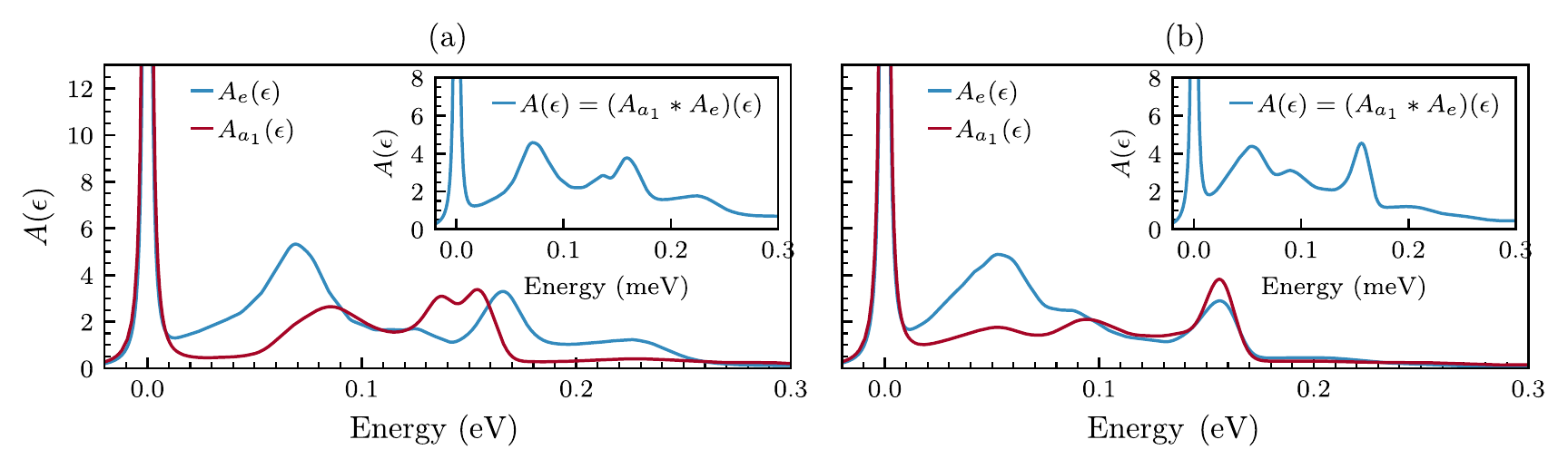}
\caption{Spectral functions $A(\epsilon)$ (in 1/eV) for optical transitions: (a) $\tA2 \to \dE$ (photoionization from the ground state $\tA2$);
(b) $\tE \to \qA2$ (photoionization from the excited state $\tE$). Blue: the contribution of $a_1$ modes, $A_{a_1}(\epsilon)$; 
red: the contribution of $e$ modes, $A_{e}(\epsilon)$. The inset shows the total spectral function $A(\epsilon)$, which is a convolution
of $a_1$ and $e$ components.
\label{fig:A}}
\end{center}
\end{figure}

The spectral function $A(\epsilon)$ describes the lineshape of the
optical transition due to coupling to lattice vibrations. The formal
definition and the {\it ab-initio\/} methodology to calculate
$A(\epsilon)$ is presented in Ref.~\cite{razinkovas2020} and briefly
outlined below.

The quantities required to compute $A(\epsilon)$ are the spectral
densities of electron--phonon coupling, defined as
$S(\hbar\omega)=\sum_{k}S_{k}\delta(\hbar\omega-\hbar\omega_k)$ \cite{razinkovas2020}.  Here
the sum runs over vibrational modes $k$ with angular frequencies
$\omega_k$, and $S_k$ are partial Huang--Rhys
factors~\cite{davies1981}.  The calculated PBE densities of
electron--phonon coupling for $a_1$ and $e$ symmetry modes are shown
in Fig.~\ref{fig:hr_spectrum}: (a) for the $\tA2 \to \dE$ transition
(photoionization from the ground state $\tA2$); (b) for the
$\tE \to \qA2$ transition (photoionization from the excited state
$\tE$). $a_2$-symmetry modes do not participate in the optical
transition. Vibrational structure was calculated in
$16\times16\times16$ supercells (with 32\,768 atomic sites) using the
embedding methodology discussed in
Ref.~\cite{razinkovas2020}. $\delta$-functions in the definition of
$S(\hbar \omega)$ have been replaced by Gaussians with width
$\sigma=5$~meV.

Both of the investigated transitions are such that either the initial
or the final state is an orbital doublet of $E$ symmetry that is
subject to a dynamical Jahn-Teller effect due to coupling to $e$
vibrations (so-called $E\otimes e$ Jahn--Teller systems).  The
contribution of Jahn--Teller active $e$ symmetry modes to the spectral
function, $A_e(\epsilon)$, was calculated using the multi-mode
Jahn--Teller treatment presented in Ref.~\cite{razinkovas2020}. This
approach involves the diagonalization of the vibronic Hamiltonian in
the basis of a selected set of effective $e$ modes. For the current
calculation we chose 20 effective modes that approximate the actual
spectral density $S_e(\hbar\omega)$ \cite{razinkovas2020}.  Obtained
spectral functions $A_e(\epsilon)$ for the two transitions are shown
in Figs.~\ref{fig:A} (a) and (b).  The contribution of $a_1$ symmetry
modes to the spectral function, $A_{a_1}(\epsilon)$, was calculated
using the generating function
approach~\cite{alkauskasNJP2014,razinkovas2020} and is also shown in
Figs.~\ref{fig:A} (a) and (b). The final spectral function
$A(\hbar \omega)$ is a convolution of the two contributions $A_e$ and
$A_{a_1}$, shown as insets in Figs.~\ref{fig:A} (a) and~(b).

\section{Brillouin-zone integration}

\begin{figure}
\begin{center}
\includegraphics[width=1\linewidth]{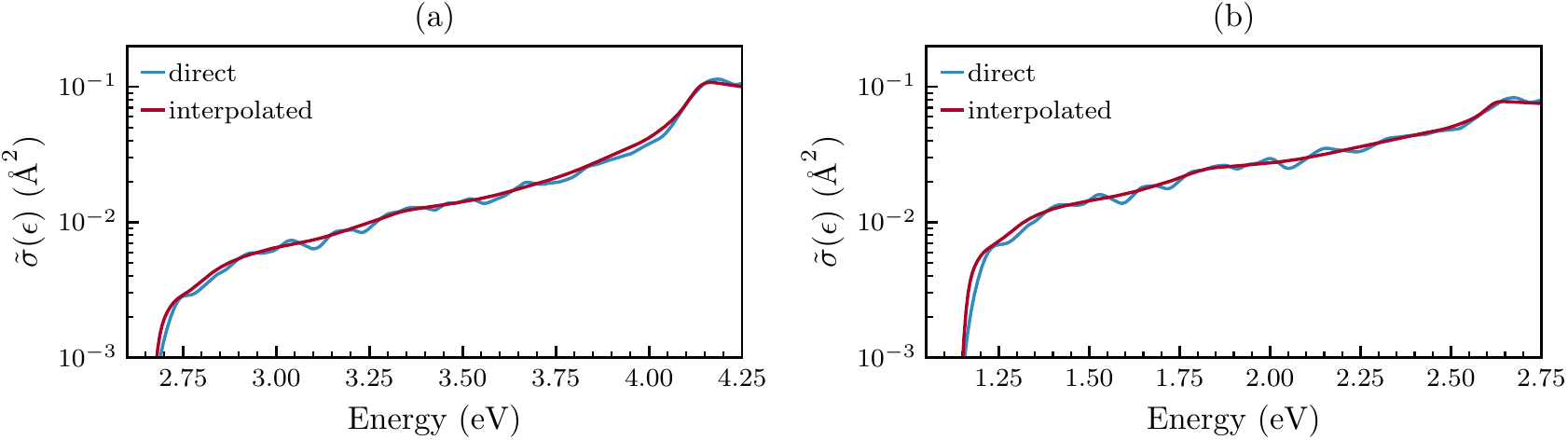}
\caption{$\tilde{\sigma}_{\text{ph}}(\epsilon)$ [Eq.~(3) of the main text] calculated via a direct summation over a $14\times 14\times 14$ $k$-point mesh (blue lines)
and using an interpolation procedure discussed in the main text (red lines). (a) Photoionization from the ground state $\tA2$; (b) photoionization
from the excited state $\tE$. \label{fig:direct_vs_interpolated}}
\end{center}
\end{figure}

The methodology for the Brillouin zone integration when calculating
$\tilde{\sigma}_{\text{ph}}(\epsilon)$ [Eq.~(3) of the main text] was
discussed in Sec.~III D of the main text. In
Fig.~\ref{fig:direct_vs_interpolated} our procedure is illustrated for
the photoionization from the ground and the excited state. Blue curves
show direct summation over a $14\times 14\times 14$ $k$-point mesh of the
$4\times4\times4$ supercell, whereby $\delta$-functions in Eq.~(3) of
the main text were replaced by Gaussians with $\sigma=20$~meV. Despite
a very dense mesh these curves display oscillations in the calculated
cross sections due to the reasons discussed in Sec.~III D of the main
text. The red curves show $\tilde{\sigma}_{\text{ph}}(\epsilon)$
evaluated by the procedure proposed in Sec.~III D of the main text.
In short, the energy of the final state is replaced with the
corresponding conduction band value taken from the bulk calculation, while
the matrix element is averaged in the Brillouin zone of the primitive
cell. The integration can be performed on a very dense $k$-point mesh.
Here we use a uniform $300\times 300\times 300$ $k$-point mesh in the Brillouin
zone of the primitive cell. This leads to a very good description
of the density of electronic states and yields smooth cross sections. 
For the final result a $10$~meV Gaussian broadening was applied.

\section{Energy-dependence of optical matrix elements}

\begin{figure}
\begin{center}
\includegraphics[width=0.95\linewidth]{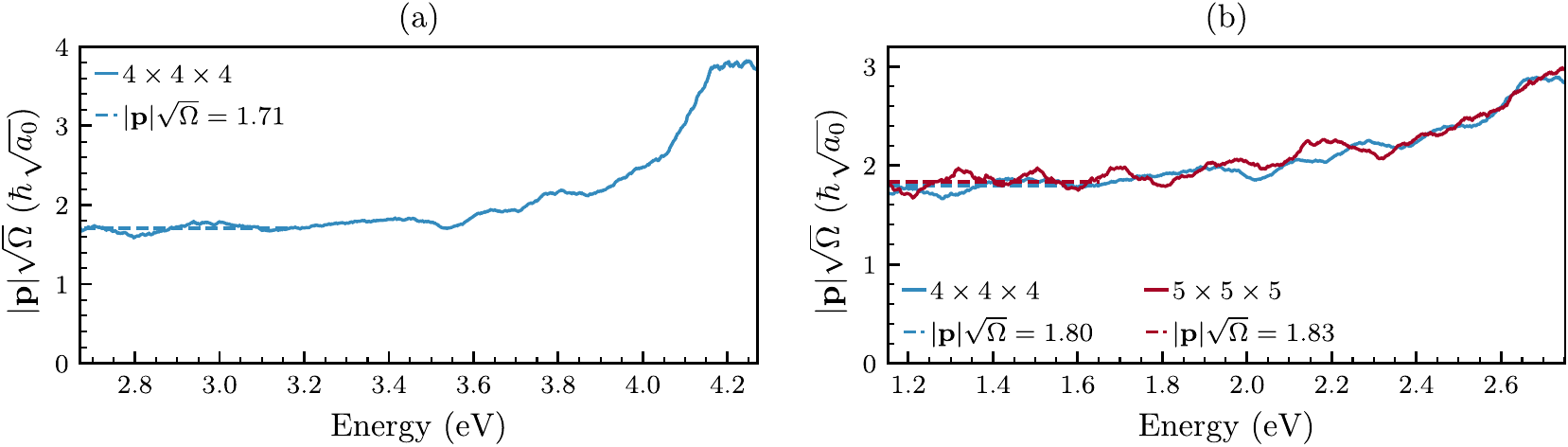}
\caption{
  Dependence of momentum matrix elements on the energy of transition for (a) ${\tA2}\to{\dE}$ (photoionization from $\tA2$) and (b)
  ${\tE}\to{\qA2}$ (photoionization from $\tE$) transitions. Calculated values are averaged by considering transitions with energies within 50~meV. 
  Blue curve: calculations for the $4 \times 4 \times 4$ supercell and a $14 \times 14 \times 14$ mesh for the $k$-point integration. 
  Red curve: calculations for the $5 \times 5 \times 5$ supercell and $8 \times 8 \times 8$ mesh for the $k$-point integration.
  Numerical values of the quantity $|\vec{p}_{ij}|\sqrt{\Omega}$ above the threshold are also indicated.
  \label{fig:p-const}}
\end{center}
\end{figure}

As discussed in Sec.~IV B of the main text, we find that momentum
matrix elements $\vec{p}_{ij}=i m (\epsilon/\hbar) \vec{r}_{ij}$ (or,
more precisely, $p^2_{ij}$) attain a constant value close to the
CBM. This is illustrated in Fig.~\ref{fig:p-const}, where we show
$|\vec{p}_{ij}|\sqrt{\Omega}$ as a function of the transition energy
(here $\Omega$ is the volume of the supercell).  Interestingly, we
find that $|\vec{p}_{ij}|$ is constant up to 0.8~eV above the
threshold both in the case of photoionization from the the ground and
the excited state. Since we deal with the optical transition to a
delocalized state, the actual value of $|\vec{p}_{ij}|$ depends on the
volume of the supercell $\Omega$. This is because delocalized
states are normalized to 1 over the volume of the supercell. However,
we find that the value $|\vec{p}_{ij}| \sqrt{\Omega}$ is already
converged within 2\% for the $4 \times 4 \times 4$ supercells used in
our work, as illustrated in Fig.~\ref{fig:p-const}(b). The actual
numerical values of $|\vec{p}_{ij}| \sqrt{\Omega}$ close to the
threshold are also indicated in Fig.~\ref{fig:p-const}.

\bibliography{references}